\documentclass[twocolumn,superscriptaddress,prd,showkeys,showpacs,nofootinbib]{revtex4-1}
\RequirePackage{lineno}\newdimen\linenumbersep\linenumbersep=2pt

\usepackage{graphicx}
\usepackage[colorlinks = true,
            linkcolor = cyan,
            urlcolor  = blue,
            citecolor = red,
            anchorcolor = blue]{hyperref}\usepackage{color}
\usepackage{amssymb}
\usepackage[nointegrals]{wasysym}
\usepackage{amsthm}
\usepackage{amsmath}
\usepackage{textcomp}
\usepackage{mathtools}
\usepackage{subcaption}
\usepackage{multirow}
\usepackage{upgreek}
\usepackage{isotope}
\usepackage{caption}
\usepackage{lineno}

\usepackage{comment}
\usepackage[separate-uncertainty,retain-explicit-plus,per-mode = symbol]{siunitx}
\usepackage{url}
\usepackage{array}
\usepackage{nicefrac}
\newcolumntype{C}[1]{>{\centering\arraybackslash}p{#1}}
\renewcommand{\arraystretch}{1.5}

\begin{document}

\title{Detection prospects for the second-order weak decays of $^{124}$Xe in multi-tonne xenon time projection chambers}
\author{Christian~Wittweg}\email[]{c.wittweg@wwu.de} \affiliation{Institut f{\"u}r Kernphysik, Westf{\"a}lische Wilhelms-Universit{\"a}t M{\"u}nster, 48149 M{\"u}nster, Germany}
\author{Brian~Lenardo} \affiliation{Physics Department, Stanford University, 382 Via Pueblo Mall, Stanford, CA 94305, USA} 
\author{Alexander~Fieguth} \affiliation{Physics Department, Stanford University, 382 Via Pueblo Mall, Stanford, CA 94305, USA}
\author{Christian~Weinheimer} \affiliation{Institut f{\"u}r Kernphysik, Westf{\"a}lische Wilhelms-Universit{\"a}t M{\"u}nster, 48149 M{\"u}nster, Germany}


\date{\today}
\begin{abstract}
We investigate the detection prospects for two-neutrino and neutrinoless second order weak decays of \isotope[124]{Xe} -- double electron capture ($0/2\upnu\text{ECEC}$), electron capture with positron emission ($0/2\upnu\text{EC}\upbeta^+$) and double-positron emission ($0/2\upnu\upbeta^+\upbeta^+$) -- in multi-tonne xenon time projection chambers. We simulate the decays in a liquid xenon medium and develop a reconstruction algorithm which uses the multi-particle coincidence in these decays to separate signal from background. This is used to compute the expected detection efficiencies as a function of position resolution and energy threshold for planned experiments. In addition, we consider an exhaustive list of possible background sources and find that they are either negligible in rate or can be greatly reduced using our topological reconstruction criteria. In particular, we draw two conclusions: First, with a half-life of $T_{1/2}^{2\upnu\text{EC}\upbeta^+} = (1.7 \pm 0.6)\cdot 10^{23}\,\text{yr}$, the $2\upnu\text{EC}\upbeta^+$ decay of $^{124}$Xe will likely be detected in upcoming Dark Matter experiments (e.g. XENONnT or LZ), and their major background will be from gamma rays in the detector construction materials. Second, searches for the $0\upnu\text{EC}\upbeta^+$ decay mode are likely to be background-free, and new parameter space may be within the reach. To this end we investigate three different scenarios of existing experimental constraints on the effective neutrino mass. The necessary 500 kg-year exposure of $^{124}$Xe could be achieved by the baseline design of the DARWIN observatory, or by extracting and using the $^{124}$Xe from the tailings of the nEXO experiment. We demonstrate how a combination of $^{124}$Xe results with those from $0\upnu\upbeta^-\upbeta^-$ searches in $^{136}$Xe could help to identify the neutrinoless decay mechanism.
\end{abstract}


\maketitle





\section{Introduction}
\label{sec:intro}
The origin of the matter-antimatter asymmetry in the Universe and the mechanism generating the neutrino masses are among the great unsolved questions of modern particle physics. Neutrinoless second-order weak decays are one of the experimental channels available to address these questions by testing the Majorana nature of neutrinos \cite{Majorana1937,Bernabeu:1983yb,Sujkowski:2003mb, buchmueller05leptogenesis}. Most experimental effort to date has focused on searching for the neutrinoless double beta decay ($0\upnu\upbeta^-\upbeta^-$) of neutron-rich candidate isotopes~\cite{KAMLANDZen_2016,EXO200CompleteExposure_2019,Majorana26kgResults_2019,GerdaLimit_2018,CUOREFirstResults_2018}, due to their relatively high natural abundance compared to proton-rich candidates. However, proton-rich isotopes offer unique decay topologies that make them of considerable experimental interest as well. In particular, those with Q-values greater than 2044~keV ($4m_e c^2$) can decay in three possible modes -- double-electron capture ($0/2\upnu\text{ECEC}$), double-positron emission ($0/2\upnu\upbeta^+\upbeta^+$) and single electron-capture with coincident positron emission ($0/2\upnu\text{EC}\upbeta^+$) \cite{dec-Doi:1991xf} -- which each produce a different experimental signature. In detectors with high-fidelity position reconstruction, tagging the specific combinations of emitted particles would be a powerful tool for discriminating signal events from backgrounds, potentially providing an extremely low-background or background-free experiment.

While searches for the neutrinoless decays can complement $0\upnu\upbeta^-\upbeta^-$ searches~\cite{Bernabeu:1983yb, Sujkowski:2003mb} positronic, second-order decays with neutrino emission are theoretically well-established~\cite{dec-Winter:1955zz}. Here, new measurements can be used as a benchmark for nuclear matrix element calculations at the long half-life extreme.

The isotope $^{124}$Xe is of particular interest as its Q-value of \SI{2856.73(12)}{\kilo\eV} \cite{dec-Nesterenko:2012xp} energetically allows all three two-neutrino and neutrinoless decay modes. Its double-K-electron capture ($2\upnu$ECEC) has recently been measured with the XENON1T Dark Matter detector~\cite{dec-XENON:2019dti}. At $T_{1/2}^{2\upnu\text{KK}} = (1.8 \pm 0.5_\text{stat} \pm 0.1_\text{sys}) \times 10^{22}\;\text{yr}$ the measurement agrees well with recent theoretical predictions \cite{dec-Suhonen:2013rca,nme-PhysRevC.91.054309,dec-Perez:2018cly}. In this decay, the measurable signal is constituted by the atomic deexcitation cascade of X-rays and Auger electrons that occurs when the vacancies of the captured electrons are refilled. In the XENON1T measurement this cascade was resolved as a single signal at 64.3 keV.  
An observation of the KL-capture and LL-capture \cite{dec-Doi:1991xf} could be within reach in future experiments if background levels can be controlled,
which would allow the decoupling of the nuclear matrix element from phase-space factors. Furthermore, the discovery potential for the positron-emitting modes (2$\upnu$EC$\upbeta^+$ or 2$\upnu\upbeta^+\upbeta^+$) in future, longer-exposure experiments could be enhanced by their distinct experimental signatures \cite{dec-Barros:2014exa}. Position-sensitive detectors could tag the $\upgamma$-rays emitted by the annihilating positron, providing a tool for rejecting $\upgamma$-ray and $\upbeta$-decay backgrounds which arise from natural radioactivity. 
In beyond-the-Standard-Model neutrinoless decays, the entire energy must be emitted in the form of charged particles or photons, favoring the positron-emitting decay channel  $0\upnu$EC$\upbeta^+$~\cite{psf-PhysRevC.87.057301, dec-PhysRevD.27.2765, dec-doi_neutrinoless, dec-Hirsch1994, dec-SUHONEN2003271, dec-PhysRevC.80.044303}. As in the two-neutrino case, the coincidence signature of the atomic relaxation, the mono-energetic positron and the two subsequent back-to-back $\upgamma$-rays could be used to reject background. $^{124}$Xe may also allow a resonant enhancement in $0\upnu$ECEC to an excited state of $^{124}$Te \cite{dec-Nesterenko:2012xp}, which would be needed to provide accessible experimentally half-lives~\cite{PhysRevC.89.064319}. The experimental signature contains multiple $\gamma$-rays emitted in a cascade, so coincidence techniques can be used to increase experimental sensitivity by suppressing the background substantially. 

Liquid xenon time projection chambers (TPCs) are ideally suited to search for \isotope[124]{Xe} decays, due to their large relatively target masses with 1~kg of \isotope[124]{Xe} per tonne of natural xenon, low backgrounds, $\mathcal{O}(\SI{1}{\%})$ energy resolution at $Q = 2.8$\,MeV, and position reconstruction for individual interactions within an event. In this work, we investigate the detection prospects of $2\upnu\text{EC}\upbeta^+$, $2\upnu\upbeta^+\upbeta^+$, $0\upnu\text{ECEC}$, $0\upnu\text{EC}\upbeta^+$ and $0\upnu\upbeta^+\upbeta^+$ in multi-tonne xenon TPCs such as the next-generation Dark-Matter detectors LZ \cite{lz-Akerib:2019fml}, PandaX-4t \cite{Zhao:2018pdr} and XENONnT \cite{aprile15c}, as well as the future nEXO \cite{Albert:2017hjq} double-$\upbeta$ decay experiment, and the DARWIN \cite{Aalbers:2016jon} Dark Matter detector. We simulate the experimental signatures of the second-order $^{124}$Xe decays in such detectors, compute the expected signal detection efficiencies, assess background sources, and calculate the experimental sensitivity as a function of the $^{124}$Xe exposure. We close with a brief discussion on the physics case for pursuing these efforts.

The signal modeling and estimated half-lives of $^{124}$Xe are discussed in \ref{sec:signal}. Relevant details of liquid xenon TPCs are described in \ref{sec:detector}. The detection efficiencies for the different decay channels will be affected by a given detector's energy resolution, spatial resolution, energy threshold and exposure. We outline the analysis of these effects and give the resulting efficiencies in section \ref{sec:analysis}. Potential backgrounds and their impact are discussed in section \ref{sec:background}. The experimental sensitivities are then given in \ref{sec:sensitivity} and followed by the discussion in section \ref{sec:discussion}.

\section{Signals from $^{124}$Xe Decay}
\label{sec:signal}
The decay modes under investigation provide distinct signatures that can be measured by the coincidence and magnitude of energy depositions (Table \ref{tab:decays}) in a detector. We group the decay modes by the number of emitted positrons. Each emitted positron will lead to the emission of at least two $\upgamma$-rays and reduce the energy that is initially available for the positrons and neutrinos by twice the positron mass. Each of the $0\upnu$ decays will exhibit a monoenergetic total energy deposition while the $2\upnu$ decays have continuous spectra due to the neutrinos leaving the detector without further interaction. 

We only consider decays to the ground state of the daughter nucleus for the positronic decay modes. A special treatment is required for $0\upnu\text{ECEC}$, as only decays which resonantly populate an excited state of \isotope[124]{Te} may be experimentally accessible.

\subsection{Signal models of decay modes}
\begin{table*}    
\centering
    \begin{center}
            \caption{Signatures of the different decay modes of \isotope[124]{Xe}.}
            \label{tab:decays}
            \begin{tabular}{c c c}
             \hline
         \textbf{Decay mode} & \textbf{Emitted quanta} & \textbf{Coincidence} \\
         \hline
             \hline
         $2\upnu\text{EC}\upbeta^+$ & X-rays/e$_\text{Auger}$, e$^+$, $2\upnu$ & X-rays/e$_\text{Auger}$ + e$^+$ and $2\upgamma$ from (e$^+$+e$^-$) \\
         $0\upnu\text{EC}\upbeta^+$ & X-rays/e$_\text{Auger}$, e$^+$ & X-rays/e$_\text{Auger}$ + e$^+$ and $2\upgamma$ from (e$^+$+e$^-$) \\
         $2\upnu\upbeta^+\upbeta^+$ & $2\text{e}^+$, $2\upnu$ & $2$e$^+$ and $4\upgamma$ from (2e$^+$+2e$^-$) \\
         $0\upnu\upbeta^+\upbeta^+$ & $2\text{e}^+$ & $2$e$^+$ and $4\upgamma$ from (2e$^+$+2e$^-$) \\
         $0\upnu\text{ECEC}$ & X-rays/e$_\text{Auger}$, $2-3\upgamma$ & X-rays/e$_\text{Auger}$ and $3\upgamma$ \\
         \hline
            \end{tabular}
    \end{center}
\end{table*}
\subsubsection{$0/2\upnu\text{EC}\upbeta^+$}
The electron capture with coincident positron emission can be written as
\begin{align}
    \isotope[124]{Xe} + \text{e}^- \rightarrow \isotope[124]{Te} + \text{e}^+ (+ 2\upnu_\text{e}) + \text{X}_\text{k},
\end{align}
where the Standard Model decay features the emission of two electron-neutrinos ($\upnu_\text{e}$) in addition to the positron ($\text{e}^+$). We assume the most-likely case of an electron capture from the K-shell. This will produce a cascade of X-rays and Auger electrons ($\text{X}_\text{k}$) with a total energy of \SI{31.8115(12)}{\kilo\eV} \cite{nist-RevModPhys.75.35}. The total available energy for the $\text{e}^+$ and the two $\upnu_\text{e}$ is then given by
\begin{align}
E_\text{e} (+E_{2\upnu}) &= Q - 2m_\text{e}c^2 - E_\text{k} \nonumber\\
                        &= \SI{2856.73(12)}{\kilo\eV}-\SI{1022.00}{\kilo\eV}-\SI{31.81}{\kilo\eV}\nonumber\\
                        &=\SI{1802.92(12)}{\kilo\eV},
\end{align}
where one has a monoenergetic positron for the neutrinoless decay and a $\upbeta$-like spectrum for the two-neutrino decay. Upon thermalization the $\text{e}^+$ annihilates with an atomic electron resulting in two back-to-back \SI{511}{\kilo\eV} $\upgamma$-rays\footnote{The electron mass uncertainty of \SI{44}{ppb} and the uncertainty on the K-shell X-ray energy in xenon are neglected in our calculations, as they will not affect the results. Moreover, we note that the $2\upgamma$-annihilation is by far the most likely case for positronium, but more $\upgamma$-rays are possible.}. 

\subsubsection{$0/2\upnu\upbeta^+\upbeta^+$}
The reaction equation for the $\upbeta^+\upbeta^+$-decay to the ground state is
\begin{align}
    \isotope[124]{Xe} \rightarrow \isotope[124]{Te} + \text{2e}^+ (+ 2\upnu_\text{e}).
\end{align}
The energy available for the two $\text{e}^+$ and the two $\upnu_\text{e}$ is given by
\begin{align}
E_{2\text{e}} (+E_{2\upnu}) &= Q - 4m_\text{e}c^2\nonumber\\
                        &= \SI{2856.73(12)}{\kilo\eV}-\SI{2043.99}{\kilo\eV}\nonumber\\
                        &=\SI{812.74(12)}{\kilo\eV},
\end{align}
where one has a continuous spectrum for the energies of the two positrons for the two-neutrino decay and a peak for the neutrinoless decay. Upon thermalization the positrons annihilate to at least four \SI{511}{\kilo\eV} $\upgamma$-rays emitted as back-to-back pairs. We do not model the angular correlation of the positrons, as their thermalization range is smaller than the spatial resolution in existing and planned experiments.

\subsubsection{Resonant $0\upnu\text{ECEC}$}
\begin{figure}[t]
    \centering
    \includegraphics[width=\linewidth]{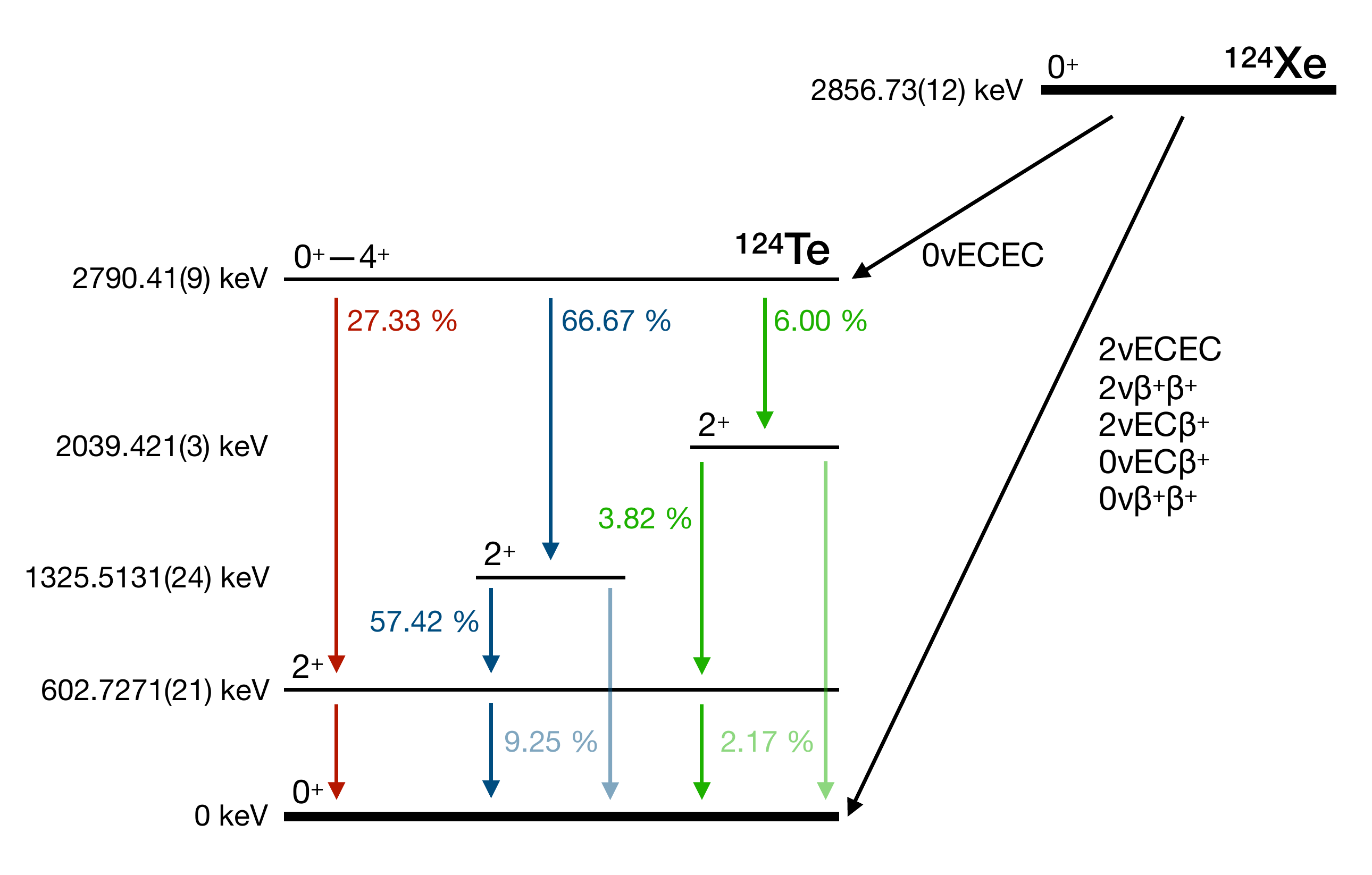}
    \caption{Decay scheme of \isotope[124]{Xe}. While $0/2\upnu\text{EC}\upbeta^+$, $0/2\upnu\upbeta^+\upbeta^+$ and $2\upnu\text{ECEC}$ most likely occur to the ground state of \isotope[124]{Te}, $0\upnu\text{ECEC}$ resonantly populates an excited state at \SI{2790.41(9)}{\kilo\eV}. There are five different known $\upgamma$-cascades along three different intermediate states. The energy level and $J^P$ are given for each state and the $\upgamma$-intensities $I_{\upgamma,\text{i}}$ for the transitions have been normalized, such that $\sum_\text{i}I_{\upgamma,\text{i}}=\SI{100}{\%}$ \cite{KATAKURA20081655}.} 
    \label{fig:decay_scheme}
\end{figure}

In contrast to the former decay modes, the energy released in the $0\upnu\text{ECEC}$ decay has to be transferred to a matching excited nuclear state $\isotope[124]{Te}^*$ of the daughter isotope, since no initial quanta are emitted from the nucleus. For a double-K capture one only has the atomic deexcitation cascade ($\text{X}_\text{2k}$): 
\begin{align}
    \isotope[124]{Xe} + 2\text{e}^- \rightarrow &\isotope[124]{Te^*}+\text{X}_\text{2k},\nonumber\\
    &\isotope[124]{Te^*} \rightarrow \isotope[124]{Te} + \;\text{multiple}\;\upgamma. 
\end{align}
The corresponding energy match has to be exact within uncertainties to avoid a violation of energy and momentum conservation. Therefore, the excitation energy $E_{\text{exc,res}}$ of the state $\isotope[124]{Te}^*$ has to fulfill the resonance condition
\begin{align}
E_{\text{exc,res}} &= Q - E_\text{2K}\nonumber\\
                        &= \SI{2856.73(12)}{\kilo\eV}-\SI{64.457(12)}{\kilo\eV}\nonumber\\
                        &=\SI{2792.27(13)}{\kilo\eV} .
\end{align}
Here, $E_\text{2k}=\SI{64.457(12)}{\kilo\eV}$ is the energy of the double electron hole after a double-K capture \cite{dec-Nesterenko:2012xp} that occurs in \SI{76.6}{\%} of all decays \cite{dec-doi_neutrinoless}. The resonance is approximately realized with a positive parity nuclear state at $E_\text{exc,res} = \SI{2790.41(9)}{\kilo\eV}$ and a corresponding deviation of \SI{1.86(15)}{\kilo\eV}\footnote{The authors of \cite{dec-Nesterenko:2012xp} recommend to perform at least one more independent measurement of the \isotope[124]{Xe}$\rightarrow$\isotope[124]{Te} Q-value in order to resolve discrepancies between existing measurements. In addition a determination of $J^P$ of the \SI{2790.41(9)}{\kilo\eV} excited state would be helpful in order to further assess the feasibility of this decay mode.} \cite{dec-Nesterenko:2012xp, KATAKURA20081655}. The angular momentum of this state is not precisely known, but $0^+$ to $4^+$ are possible $J^P$ configurations. The level scheme relevant to the decay is shown in shown in Fig.~\ref{fig:decay_scheme}. There are five different $\upgamma$-cascades that are either $\geq 0^+ \rightarrow 2^+ \rightarrow 0^+$ or $\geq 0^+ \rightarrow 2^+ \rightarrow 2^+ \rightarrow 0^+$ for two- and three-$\upgamma$ transitions, respectively. As a considerable decay rate is only expected to $0^+$ and $1^+$ states \cite{dec-Nesterenko:2012xp}, we assume that the resonantly populated state is $0^+$ and focus on the $0^+ \rightarrow 2^+ \rightarrow 2^+ \rightarrow 0^+$ transition that occurs in \SI{57.42}{\%} of all decays.

\subsection{Half-life calculations}

\subsubsection{Two-neutrino decays}
The half-life predictions for the two-neutrino decay modes are constructed from
\begin{equation}
    (T^{2\upnu}_{1/2})^{-1} = G_{2\upnu} |M_{2\upnu}|^2,
    \label{ref:nme2nu}
\end{equation}
where $G_{2\upnu}$ is a phase-space factor (PSF) and $|M_{2\upnu}|^2$ is the nuclear matrix element (NME). While the PSF is different among the decay modes \cite{dec-Doi:1991xf, psf-PhysRevC.87.057301,psf-Stoica:2019ajg}, the NME differs only slightly between 2$\upnu$ECEC and 2$\upnu$EC$\upbeta^+$ and is about a factor of two smaller for $2\upnu\upbeta^+\upbeta^+$ \cite{nme-PhysRevC.91.054309, dec-Suhonen:2013rca}. For simplicity, we assume 
\begin{align}
\label{eq:NME_relation_2nu}
M_{2\upnu\text{ECEC}}=M_{2\upnu\text{EC}\upbeta^+}=2\times M_{2\upnu\upbeta^+\upbeta^+}
\end{align}
and use the existing ${2\upnu\text{ECEC}}$ measurement to constrain $M_{2\upnu\text{ECEC}}$. This is justified by the relatively large uncertainty from the measured half-life \cite{dec-XENON:2019dti} which outweighs the expected NME differences. 

As only the value for the double K-capture has been reported, the half-life has to be scaled by the fraction of double-K decays $f^{2\upnu\text{KK}} = 0.767$ \cite{dec-doi_neutrinoless}. One obtains a total half-life of
\begin{align}
\label{eq:dec_xe124_all}
T^{2\upnu\text{ECEC}}_{1/2}&=(1.4\pm 0.4)\times 10^{22}\;\text{yr}.
\end{align}
Using Eq.~(\ref{ref:nme2nu}) with the measured half-life and calculated PSFs one has
\begin{align}
   \label{eq:half-life_relation_2nu}
    T^{2\upnu\text{EC}\upbeta^+}_{1/2} &= \frac{G_{2\upnu\text{ECEC}}}{G_{2\upnu\text{EC}\upbeta^+}} \times T^{2\upnu\text{ECEC}}_{1/2},\nonumber\\
    T^{2\upnu\upbeta^+\upbeta^+}_{1/2} &= \frac{4 \times G_{2\upnu\text{ECEC}}}{G_{2\upnu\upbeta^+\upbeta^+}} \times T^{2\upnu\text{ECEC}}_{1/2}.
\end{align}{}
The resulting expected half-lives for $2\upnu\text{EC}\upbeta^+$ and $2\upnu\upbeta^+\upbeta^+$ are given in Tab. \ref{tab:half-lives-2nu}. Due to the smaller available phase-spaces,
the $2\upnu\text{EC}\upbeta^+$ half-life is about one order of magnitude longer than the one for $2\upnu\text{ECEC}$, whereas the $2\upnu\upbeta^+\upbeta^+$ half-life is about six orders of magnitude longer. This makes $2\upnu\text{EC}\upbeta^+$ a promising target for next-generation experiments such as LZ or XENONnT while the double-positronic mode will be challenging to measure.

\begin{table*}[]
    \centering
    \begin{center}
            \caption{The different 2$\upnu$ decay modes of \isotope[124]{Xe} with the corresponding phase-space factors (PSF), the assumptions of the corresponding matrix elements according to Eq.\,(\ref{eq:NME_relation_2nu}), and the measured or predicted half-lives according to Eq.\,(\ref{eq:dec_xe124_all}) and Eq.\,(\ref{eq:half-life_relation_2nu}), respectively. 
            The PSF values were taken from the review \cite{psf-Stoica:2019ajg} which summarizes 
            work by the reviewers and from  \cite{dec-PhysRevC.87.024313,dec-Doi:1991xf}. 
            Therefore, we give a range of PSF values. For predicting the half-lives of the decay modes ${2\upnu\text{EC}\upbeta^+}$ and ${2\upnu\upbeta^+\upbeta^+}$ we use the central value of this range as the most probable PSF value and half of this range as the uncertainty.} 
            \label{tab:half-lives-2nu}
            \setlength{\tabcolsep}{8pt}
            \begin{tabular}{ c c c c c }
             \hline
       \textbf{Decay mode} & $\mathbf{G_{2\upnu}}$ \textbf{[yr$\mathbf{^{-1}}$]} & $\mathbf{M_{2\upnu}}$ 
       (\ref{eq:NME_relation_2nu}) & \multicolumn{2}{c}{\textbf{Half-life [yr]}} \\     
      & & & \textbf{Measured (\ref{eq:dec_xe124_all})} & \textbf{Predicted (\ref{eq:half-life_relation_2nu})}\\
         \hline
       ${2\upnu\text{ECEC}}$ & $(1.5-2.0)\cdot 10^{-20}$ & $M_{2\upnu\text{ECEC}}$ & $(1.4 \pm 0.4) \cdot 10^{22}$ & \\
        ${2\upnu\text{EC}\upbeta^+}$ & $ (1.2-1.7)\cdot 10^{-21}$ & $ M_{2\upnu\text{ECEC}}$ & & $(1.7 \pm 0.6) \cdot 10^{23}$ \\
         ${2\upnu\upbeta^+\upbeta^+}$ & $(4.3-4.9)\cdot 10^{-26}$ & $\frac{1}{2}\cdot M_{2\upnu\text{ECEC}}$ & & $(2.2 \pm 0.7) \cdot 10^{28}$ \\ 
         \hline 
            \end{tabular}
    \end{center}
\end{table*}


\subsubsection{Neutrinoless decays}
In case of the neutrinoless decays the equation relating PSF and NME to the half-life changes to
\begin{equation}
    (T^{0\upnu}_{1/2})^{-1} = G_{0\upnu} |M_{0\upnu}|^2 |f(m_{\text{i}},U_{\text{ei}})|^2. \label{eq:0vdecayrate}
\end{equation}
Note that the PSF ($G_{0\upnu}$) and NME ($|M_{0\upnu}|^2$) are different from those used previously due to the absence of neutrino emission.
The additional factor $f(m_{\text{i}},U_{\text{ei}})$ contains physics beyond the Standard Model. Typically the decay is assumed to proceed via light neutrino exchange, for which we have
\begin{equation}
    f(m_{\text{i}},U_{\text{ei}}) = \frac{\langle m_\upnu\rangle}{m_\text{e}} = \frac{\sum_{\text{k=light}}(U_\text{ek}^2m_\text{k})}{m_\text{e}}.
\end{equation}
Here the effective neutrino mass $\langle m_\upnu \rangle$ is a linear combination of neutrino masses $m_\text{i}$ and elements of the PMNS mixing matrix $U_{\text{ei}}$ \cite{dec-PhysRevC.87.024313, PhysRevC.89.064319}. 
For $0\upnu\text{ECEC}$ a resonance factor $R$ has to be added to Eq.\,(\ref{eq:0vdecayrate}):
\begin{equation}
    (T^{0\upnu\text{ECEC}}_{1/2})^{-1} = G_{0\upnu} |M_{0\upnu}|^2 |f(m_{\text{i}},U_{\text{ei}})|^2 \cdot R\,. \label{eq:0vececdecayrate}
\end{equation}
The mismatch $\Delta=|Q-E_\text{2k}-E_\text{exc}|=\SI{1.86(15)}{\kilo\eV}$ between the available energy and the energy level of the daughter nucleus in the excited state $E_\text{exc}$ \cite{dec-Nesterenko:2012xp} defines the resonance factor $R$, which -- with the two-hole width $\Gamma=\SI{0.0198}{\kilo\eV}$ \cite{PhysRevC.89.064319}  -- amounts to
\begin{align}
  R = \frac{m_\text{e}c^2\Gamma}{\Delta^2+\nicefrac{\Gamma^2}{4}} = 2.92 \pm 0.47\,. 
\end{align}

We take the PSF values again from the review \cite{psf-Stoica:2019ajg} which summarizes work by the reviewers and from  \cite{dec-PhysRevC.87.024313,dec-doi_neutrinoless}. In order to calculate half-life expectations for neutrinoless decays of \isotope[124]{Xe}, we also need estimates for the NME, and the effective neutrino mass $\langle m_\upnu\rangle$. The NMEs have never been measured for the neutrinoless case. Only for the case with two neutrinos a few half-lives have been determined experimentally. Unfortunately the NMEs for the $2\upnu$ and the $0\upnu$ cases are not strongly connected. Moreover, the effective neutrino mass has never been measured, and we must choose among different experimental constraints accordingly.
To account for these two sources of unknowns we use the following two approaches to get lower limits for the expected half-lives of neutrinoless double-weak decays of \isotope[124]{Xe}.

{\bf Method 1:} In the first approach, to constrain the effective neutrino mass we take the newest result from the neutrino mass experiment KATRIN which set the most stringent direct, model-independent limit on $m_\upnu<1.1$\,eV~(90\% C.L.) \cite{Aker:2019uuj}. We then combine this limit with a global fit to neutrino oscillation results \cite{Esteban:2018azc} (Fig. 11 therein). This yields an upper limit range -- corresponding to the uncertainties in the Majorana and CP-phases of the PMNS neutrino mixing matrix -- on the effective neutrino mass:
\begin{align}
    \langle m_\upnu\rangle < 0.3 - 1.1\,\text{eV}/c^2 \quad (90\%~\text{C.L.}).
    \label{eq:mnu_eff}
\end{align}
For the NMEs in our first approach, we take three available sets of calculations into account. The first set is based on the quasi-random phase approximation (QRPA) and was calculated in \cite{dec-Suhonen:2013rca}. The second comes from the interacting boson model (IBM)  \cite{PhysRevC.89.064319, PhysRevC.91.034304}. The third set is based on nuclear shell model (NSM) calculations as performed for the two-neutrino case \cite{dec-Perez:2018cly} and is limited by lower and upper values of the full shell model similar to normal neutrinoless double-$\upbeta$ decay as shown in \cite{RevModPhys.77.427} and \cite{PhysRevLett.100.052503}. Both the QRPA and NSM calculations provided good predictions of $T_{1/2}$ for 2$\upnu$ECEC while there were no $2\upnu$-predictions for IBM. We summarize the relevant PSF- and NME-values and the corresponding lower half-life limits in Tab.~\ref{tab:half-lives_0nu}. 

{\bf Method 2:} In our second approach, we use a similar idea as for the prediction of the half-lives in the 2$\upnu$ case. Instead of one measured half-life value we take the half-life limits obtained in the search for $0\upnu\upbeta^-\upbeta^-$ decay of \isotope[136]{Xe} \cite{KAMLANDZen_2016,EXO200CompleteExposure_2019}. The most stringent lower limit on the half-life comes from the KamLAND-Zen experiment \cite{KAMLANDZen_2016} with
\begin{align}
    T^{0\upnu \upbeta^-\upbeta^-,136}_{1/2} > 1.07 \cdot 10^{26}~\text{yr} \quad \text{(90\%~C.L.)}.
\end{align}
Unlike the case for the various 2$\upnu$-decays in Eq.\,(\ref{eq:half-life_relation_2nu}), the NMEs of $0\upnu$-decays of \isotope[124]{Xe} are different from the NMEs of the $0\upnu\upbeta^-\upbeta^-$-decay of \isotope[136]{Xe} and do not cancel. But for this comparison of the half-life limits of neutron-poor and neutron-rich isotopes of the same element xenon, the substantial uncertainties connected to these calculations drop out to a large extent if the NMEs are calculated within the same framework and if the main uncertainties stem from the unknown quenching $q$ of the axial coupling constant $g_\text{A}$, which can be factorized out of the NME $M$: 
\begin{align}
   \tilde{M} := \frac{M}{q\cdot g^2_\text{A}}\,
\end{align}
This is of advantage, as the uncertainties of the NME are often summarized in the $g_\text{A}$-quenching, and the quenching factors $q$ can be assumed to be similar for neutron-poor and neutron-rich nuclei of the same element \cite{achim_schwenk_private_2019}. We perform the calculations using the NMEs from the interacting boson model (IBM) \cite{PhysRevC.89.064319, PhysRevC.91.034304} which possess their main uncertainties in the quenching of the axial coupling constant. If we solve equation (\ref{eq:0vdecayrate}) or (\ref{eq:0vececdecayrate}) according to the factor $f(m_{\text{i}},U_{\text{ei}})$  and equate for two decays of different xenon istopes, we obtain:
\begin{align}
     T^{0\upnu \upbeta^+ \upbeta^+,124}_{1/2} & =  T^{0\upnu \upbeta^-\upbeta^-,136}_{1/2} \cdot \frac{G^{136}_{0\upnu \upbeta^-\upbeta^-}}{G^{124}_{0\upnu \upbeta^+\upbeta^+}} 
        \cdot \frac{|\tilde M^{136}_{0\upnu \upbeta^-\upbeta^-}|^2}{|\tilde M^{124}_{0\upnu \upbeta^+\upbeta^+}|^2}, \nonumber\\
     T^{0\upnu \text{EC} \upbeta^+,124}_{1/2} & =  T^{0\upnu \upbeta^-\upbeta^-,136}_{1/2} \cdot \frac{G^{136}_{0\upnu \upbeta^-\upbeta^-}}{G^{124}_{0\upnu \text{EC} \upbeta^+}} 
        \cdot \frac{|\tilde M^{136}_{0\upnu \upbeta^-\upbeta^-}|^2}{|\tilde M^{124}_{0\upnu \text{EC} \upbeta^+}|^2}, \nonumber\\
     T^{0\upnu \text{ECEC},124}_{1/2} & =   \frac{T^{0\upnu \upbeta^-\upbeta^-,136}_{1/2}}{R} \cdot \frac{G^{136}_{0\upnu \upbeta^-\upbeta^-}}{G^{124}_{0\upnu \text{ECEC}}} 
        \cdot \frac{|\tilde M^{136}_{0\upnu \upbeta^-\upbeta^-}|^2}{|\tilde M^{124}_{0\upnu \text{ECEC}}|^2}. \label{eq:half-life_relation_0nu} 
\end{align}
In Eq.\,(\ref{eq:half-life_relation_0nu}) the effective neutrino mass and its uncertainty drop out of the equation. We would like to underline the validity of this kind of comparison by mentioning that a similar approach has been used in Ref.~\cite{Auger:2012ar} (Fig. 6 therein) to relate effective neutrino mass limits for two different isotopes within the same theoretical framework of NME calculations. Again, we summarize the results in Tab.~\ref{tab:half-lives_relation_0nu}. 
\begin{table*}[t]
    \centering
    \begin{center}
            \caption{Predicted lower limits on the half-life of the 0$\upnu$ decay modes of \isotope[124]{Xe} according to Eq.\,(\ref{eq:0vdecayrate}) as well as Eq.\,(\ref{eq:0vececdecayrate}), \textbf{Method 1:} Limits are given for the range of $\langle m_\upnu\rangle$ from Eq.\,(\ref{eq:mnu_eff}) \cite{Aker:2019uuj,Esteban:2018azc}. The PSFs ($G_{0\upnu}$) were taken from \cite{PhysRevC.89.064319}, and the review \cite{psf-Stoica:2019ajg} which summarizes work by the reviewers and from  \cite{dec-doi_neutrinoless, psf-PhysRevC.87.057301}. We use the central value of the PSF-range as the most probable value and half of this range as the uncertainty. The same is done for the NMEs ($M_{0\upnu}$) in all cases were a range of values is given in the original publication. For $0\upnu\text{ECEC}$ the NMEs values from the quasi-random phase approximation (QRPA) \cite{dec-Suhonen:2013rca} and the interacting boson model (IBM) \cite{PhysRevC.89.064319} were used. The NME for IBM is obtained by taking the single value given in the publication and assuming $g_A = 1.269$. The NME-range for QRPA stems from the smallest and largest NME value for $g_A = 1.25$ under the assumption of different bases and short-range correlations. 
            For the $0\upnu\text{EC}\upbeta^+$ and the $0\upnu\upbeta^+\upbeta^+$ QRPA \cite{dec-Suhonen:2013rca}, NSM (calculated \cite{javier_m_private_2019} as in \cite{dec-Perez:2018cly}), and IBM \cite{PhysRevC.91.034304} NMEs were considered. The range of NMEs for QRPA and the value for IBM are obtained as above. However, for the latter an uncertainty is given in the publication instead of a value range. For the NSM the NME-range is given by different model configurations and the most probable value and uncertainty are derived in the same fashion as for QRPA. All uncertainties are propagated by drawing $10^6$ independent samples from the parameter distributions and multiplying with the upper limit on $\langle m_\upnu \rangle $. Then the $90\,\%\,\text{C.L.}$ upper limit on $T^{-1}_{1/2}$ is determined from the resulting distribution and inverted to obtain the corresponding lower half-life limit.}
            \label{tab:half-lives_0nu}
            \setlength{\tabcolsep}{8pt}
            \begin{tabular}{ c c c c c c}
             \hline
 \textbf{Decay mode} & $\mathbf{G_{0\upnu}}$ \textbf{ [yr$\mathbf{^{-1}}$]} & $\mathbf{M_{0\upnu}}$ & \textbf{Model} & \multicolumn{2}{c}{\textbf{Predicted lower $\mathbf{T_{1/2}}$ limit [yr]~(90\%~C.L.)}} \\
 &&&& $\mathbf{\langle m_\upnu\rangle < 1.1\,\textbf{eV}/c^2}$ & $\mathbf{\langle m_\upnu\rangle < 0.3\,\textbf{eV}/c^2}$\\
         \hline
         ${0\upnu\text{ECEC}}$ & $2.6 \cdot 10^{-19}$ & $1.080-1.298$ & QRPA & $1.8 \cdot 10^{29}$ & $2.4 \cdot 10^{30}$
         \\ 
         && $0.478$ & IBM & $1.2 \cdot 10^{30}$ & $1.6 \cdot 10^{31}$\\
         ${0\upnu\text{EC}\upbeta^+}$ &  $(1.7-2.3)\cdot 10^{-17}$ & $4.692-6.617$ & QRPA & $8.7 \cdot 10^{25}$ & $1.2 \cdot 10^{27}$\\ 
         && $7.63(1.23)$ & IBM & $4.8 \cdot 10^{25}$ & $6.5 \cdot 10^{26}$\\
         && $2.22-4.77$ & NSM & $1.6 \cdot 10^{26}$ & $2.2 \cdot 10^{27}$\\

         ${0\upnu\upbeta^+\upbeta^+}$ &  $(1.1-1.2)\cdot 10^{-18}$ & $4.692-6.617$ & QRPA  &  $1.5 \cdot 10^{27}$ & $2.1 \cdot 10^{28}$\\ 
         && $7.63(1.23)$ & IBM & $8.6 \cdot 10^{26}$ & $1.2 \cdot 10^{28}$\\
         && $2.22-4.77$ & NSM & $2.9 \cdot 10^{27}$ & $3.9 \cdot 10^{28}$\\

         \hline

        \hline
            \end{tabular}
    \end{center}
\end{table*}

\begin{table*}[t]
    \centering
    \begin{center}
            \caption{Predicted lower limits on the half-life of the 0$\upnu$ decay modes of \isotope[124]{Xe} according to Eq.\,(\ref{eq:half-life_relation_0nu}), \textbf{Method 2:} Limits are given for $T_{1/2}^{0\upnu \upbeta^-\upbeta^-,136} > 1.07 \cdot 10^{26}\,\text{yr}$ as measured by KamLAND Zen \cite{KAMLANDZen_2016}. The PSFs for \isotope[124]{Xe} ($G_{0\upnu}^{^{124}\text{Xe}}$) were taken from \cite{PhysRevC.89.064319}, and the review \cite{psf-Stoica:2019ajg} which summarizes work by the reviewers and from  \cite{dec-doi_neutrinoless, psf-PhysRevC.87.057301}. Those for \isotope[136]{Xe} ($G_{0\upnu}^{^{136}\text{Xe}}$) were also taken from \cite{psf-Stoica:2019ajg} and \cite{psf-PhysRevC.85.034316}, cited therein. We use the central value of the PSF-range as the most probable value and half of this range as the uncertainty. For the NMEs of \isotope[124]{Xe} ($\Tilde{M}_{0\upnu}^{^{124}Xe}$) and \isotope[136]{Xe} ($\Tilde{M}_{0\upnu}^{^{136}Xe}$) we only consider the interacting boson model (IBM) and do not multiply the NMEs with $g_\text{A}^2$. We consider \cite{PhysRevC.91.034304} for $\Tilde{M}_{0\upnu}^{^{136}Xe}$. For $0\upnu\text{ECEC}$ the NME was taken from \cite{PhysRevC.89.064319}. For $0\upnu\text{EC}\upbeta^+$ and $0\upnu\upbeta^+\upbeta^+$ the NMEs were taken from \cite{PhysRevC.91.034304}. An uncertainty on the NME is only given in this publication. All uncertainties are propagated by drawing $10^6$ independent samples from the parameter distributions and multiplying with the lower limit on $T_{1/2}^{0\upnu \upbeta^-\upbeta^-,136} $. Then the $90\,\%\,\text{C.L.}$ lower limit of the half-life is determined from the resulting distribution.}
            \label{tab:half-lives_relation_0nu}
            \setlength{\tabcolsep}{8pt}
            \begin{tabular}{ c c c c c c}
             \hline
 \textbf{Decay mode} & $\mathbf{G_{0\upnu}^{^{124}\textbf{Xe}}}$ \textbf{[yr$\mathbf{^{-1}}$]} & $\mathbf{\Tilde{M}_{0\upnu}^{^{124}\textbf{Xe}}}$ & $\mathbf{G_{0\upnu}^{^{136}\textbf{Xe}}}$ \textbf{[yr$\mathbf{^{-1}}$]} &  $\mathbf{\Tilde{M}_{0\upnu}^{^{136}\textbf{Xe}}}$ & \textbf{Predicted lower $\mathbf{T_{1/2}}$} \\
 &&&&& \textbf{limit [yr]~(90\%~C.L.)}\\
         \hline
         ${0\upnu\text{ECEC}}$ & $2.6 \cdot 10^{-19}$ & $0.297$ & $(14.54-14.58)\cdot 10^{-15}$ & 3.05 & $2.1 \cdot 10^{32}$ 
         \\ 

         ${0\upnu\text{EC}\upbeta^+}$ &  $(1.7-2.3) \cdot 10^{-17}$ & $4.74(0.76)$ & $(14.54-14.58)\cdot 10^{-15}$ & 3.05 & $8.4 \cdot 10^{27}$ \\ 

         ${0\upnu\upbeta^+\upbeta^+}$ &  $(1.1-1.2) \cdot 10^{-18}$ & $4.74(0.76)$ & $(14.54-14.58)\cdot 10^{-15}$ & 3.05 &  $1.5 \cdot 10^{29}$ \\

         \hline

        \hline
            \end{tabular}
    \end{center}
\end{table*}

Among the neutrinoless decays, the $0\upnu\text{EC}\upbeta^+$ mode is expected to have the shortest -- and thus most experimentally accessible -- half-life. The other decay modes, $0\upnu\upbeta^+\upbeta^+$ and resonant $0\upnu\text{ECEC}$, exhibit considerably longer half-lives owed to unfavourable phase-space and a lack of resonance enhancement $R$, respectively. We also note that the half-life limits calculated with the first method are systematically lower than the ones for the second. This is in line with the smaller predicted upper limits on the effective neutrino mass 
\begin{equation}
    \langle m_\upnu \rangle < 0.061 - 0.165\,\text{eV}/c^2
\end{equation}{}
given by KamLAND Zen \cite{KAMLANDZen_2016}. 

The ability to observe any of the described decay channels is given not only by the theoretical prediction on their half-lives, but also by the detection efficiencies in a given experiment. In the following sections, we discuss the detection prospects of the neutrinoless decay modes in future experiments which could have significantly larger samples of $^{124}$Xe than current detectors, such as DARWIN or nEXO. 

\section{Detector Response}
\label{sec:detector}
In a liquid xenon TPC, the energy and position of an energy deposit is reconstructed using two observed signals: scintillation light and ionization charge~\cite{doke-RevModPhys.82.2053}. The former is typically detected directly using UV-sensitive photodetectors, producing a prompt signal referred to as S1. The latter is detected by applying an electric field across the liquid xenon volume and drifting the charges to a collection plane. The charge can be either detected directly using charge-sensitive amplifiers, or extracted into a gas-phase region and accelerated, producing proportional electroluminescence light that is detected in the photodetectors. The delayed secondary signal produced by the drifted charge is referred to as S2. The combination of the two signals allows one to reconstruct the 3D-position of the interaction inside the detector: the S2 hit pattern on the collection plane gives the x-y coordinate and the S1-S2 time delay gives the depth z. The deposited energy is reconstructed using the magnitude of the S1 and S2 signals. A linear combination of both signals has been shown to greatly improve the energy resolution compared to either signal individually, due to recombination of electron-ion pairs producing anticorrelated fluctuations in the energy partitioning between light and charge~\cite{Conti:2003av,AprileAntiCorr2007}.

\begin{figure}[t]
    \centering
    \includegraphics[width=0.9\linewidth]{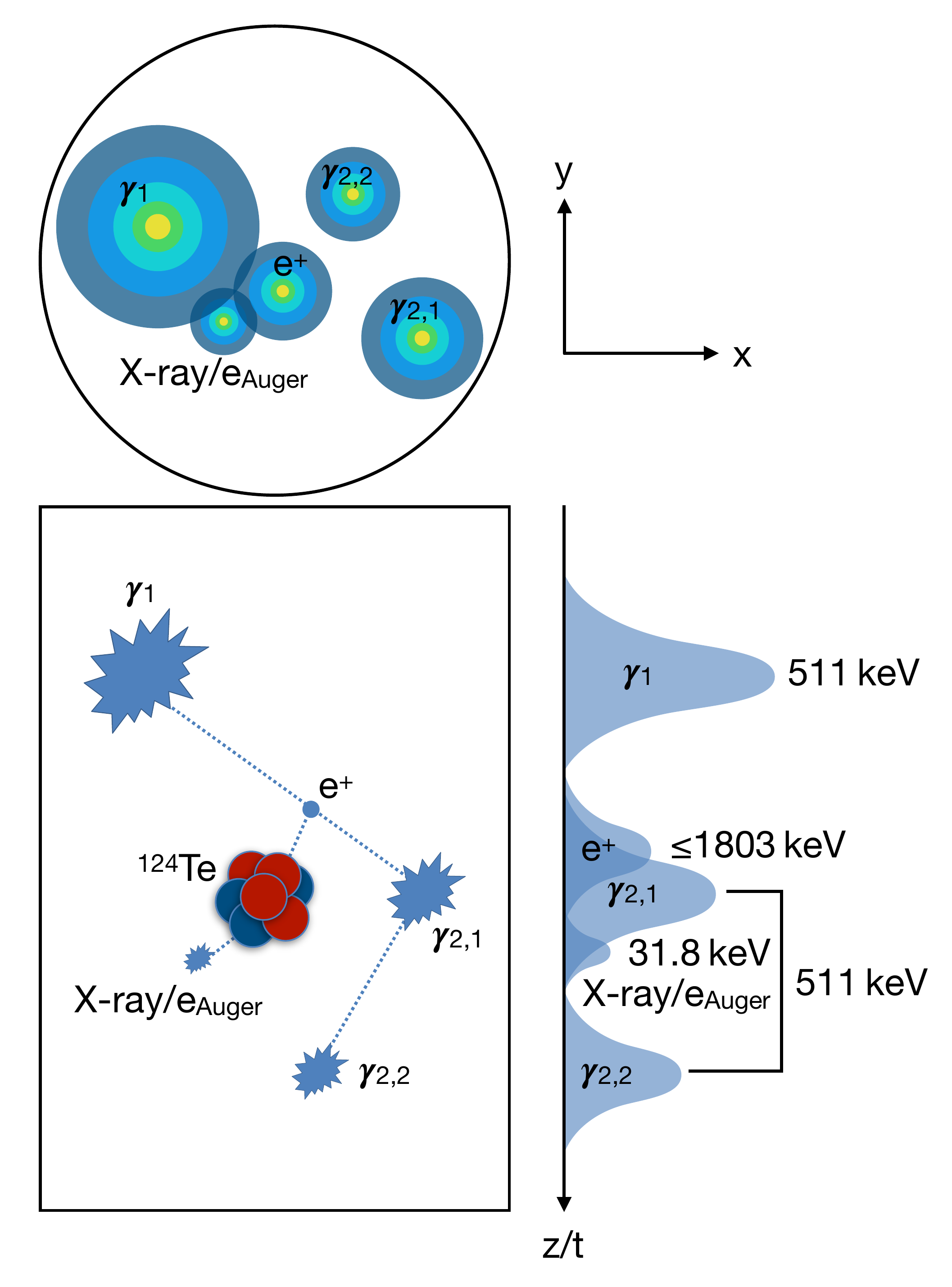}
    \caption{Schematic of a $0/2\upnu\text{EC}\upbeta^+$-decay signature inside a xenon time projection chamber. As shown in the bottom panel, an initial positron and atomic excitation quanta are emitted and deposit their energy close to the nucleus. Two secondary $\upgamma$s are emitted after the annihilation of the positron. One of those is directly absorbed and the other Compton-scatters before photoabsorption. On the $dt$/z-axis the ionization signals of $\upgamma_1$, $\upgamma_{2,1}$ and $\upgamma_{2,2}$ can be distinguished from one another. The positron and atomic deexcitation signals are merged with $\upgamma_{2,1}$ in this example. The top panel shows the corresponding hit-pattern of the ionization signal. In x-y-coordinates the individual scatters of $\upgamma_2$ can be clearly distinguished from $\upgamma_1$, while the discrimination of the atomic deexcitation quanta and the positron from $\upgamma_1$ is not trivial. The scintillation signal is merged for all energy depositions and is not shown in the figure. The sizes of the scintillation signals in x-y and z-coordinates roughly correspond to the magnitudes of the energy depositions.}
    \label{fig:analysis}
\end{figure}

For events with multiple energy deposits, the prompt S1 signals for each vertex are typically merged, resulting in a single scintillation pulse for the entire event. However, individual vertices can be resolved as individual S2 signals arriving at different positions and times on the charge collection plane. A schematic of the signature expected from a typical $0/2\upnu\text{EC}\upbeta^+$ decay of $^{124}$Xe is shown in Fig. \ref{fig:analysis}. In this example, there are five different S2 signals produced from the positron, X-ray cascade, and each of the annihilation $\gamma$-rays (one of which undergoes Compton scattering before being absorbed). With sufficient position and energy resolution, one can use this information to classify events and perform particle identification, providing a tool for separating backgrounds from the signal of interest.

The capability for a detector to resolve each vertex depends on the time resolution in the charge channel, the width of the S2 signals, and the x-y resolution of the charge collection plane. These properties are highly dependent on the specific readout techniques employed in each experiment. In addition, the detection of each energy deposit requires its individual S2 signal to be above the detector's charge energy threshold, a property that is again specific to each experiment. 

In this work, we compute the detection efficiency ($\epsilon$) for the various modes of $^{124}$Xe decay as a function of the x-y- and z-position resolution and energy threshold, to provide estimates that apply across the possible range of existing and future experiments. 

\section{Analysis}
\label{sec:analysis}
\subsection{Simulation}

We generate the emitted quanta and their initial momentum vectors for each decay channel with the event generator DECAY0 \cite{dec-tretyak}. The version used here has been modified previously for the simulation of the positronic \isotope[124]{Xe} decay modes \cite{dec-Barros:2014exa}. In the scope of this work, we verified the implementation, added the resonant $0\upnu\text{ECEC}$ decay mode, and implemented the angular correlations for the $\upgamma$-cascades under the assumption of $J^P=0^+$ for the resonantly populated state \cite{YAMAZAKI19671, Smith_2019}. In order to investigate the efficiency, at least 10$^4$ events per decay channel have been used. 

The particles generated for each decay are propagated through simplified models of the detectors under investigation using the XeSim package~\cite{la_2019}, based on Geant4 \cite{geant4-AGOSTINELLI2003250}. These detector models consist of a cylindrical liquid xenon volume in which we uniformly generate $^{124}$Xe decay events. This volume is surrounded by a thin shell of copper which is used for modeling the impact of external $\upgamma$-backgrounds. We simulate two different sizes of cylinders in this work, characteristic of two classes of future experiments. The ``Generation 2" (G2) experiments are defined as experiments which have height/diameter dimensions of between one and two meters. This class includes the LZ ~\cite{lz-Akerib:2019fml} and XENONnT Dark Matter experiments, which will use dual-phase TPCs filled with natural xenon. It also includes the future nEXO neutrinoless double-$\upbeta$ decay experiment, which will use a single-phase liquid TPC filled with xenon enriched to 90\% in $^{136}$Xe. For simplicity, we model all G2 experiments as a right-cylinder of liquid xenon with a height and diameter of 120~cm each.\footnote{LZ and XENONnT are designed slightly larger than this, but we show below that this assumption has a minimal effect on the calculated efficiency.} We also simulate a ``Generation 3" (G3) experiment, which is intended to model the proposed DARWIN Dark Matter experiment~\cite{Aalbers:2016jon}. This detector is modeled as a right-cylinder of liquid xenon with a height and diameter of 250~cm each.

For experiments using $^{\text{nat}}$Xe targets, there will be approximately 1\,kg of $^{124}$Xe per tonne of target material. The G2 Dark Matter experiments would therefore be able to reach \isotope[124]{Xe}-exposures of $\sim 50-100$\,kg-year in 10 years of run time. By scaling the target mass up to 50~tonnes, the G3 experiment DARWIN will amass an exposure of $\sim$500\,kg-year. For nEXO, the enrichment of the target in $^{136}$Xe will remove all of the $^{124}$Xe; however, here we consider the possibility of extracting the $^{124}$Xe from the depleted xenon and mixing it back into the target. There will be approximately 50\,kg of $^{124}$Xe in the nEXO tailings, meaning a 10 year experiment could amass an exposure of $\sim$500\,kg-year, competitive with a G3 natural xenon experiment.

\subsection{Energy resolution model}

Within this study all simulated detectors use the energy dependence of the resolution on the combined signal as reported in \cite{dec-XENON:2019dti}, which is modeled as
\begin{equation}
\frac{\sigma_E}{E} = \frac{a}{\sqrt{E}} + b.
\end{equation}
Here $E$ is the energy and $a=31$\,keV$^{\nicefrac{1}{2}}$ and $b=0.37$ are constants extracted from a fit to calibration data from $41-511$\,keV. The model predicts a resolution of $\sim 1\,\%$ at the Q-value of the decay, approximately consistent with the energy resolution published by the EXO-200 experiment at a similar energy~\cite{EXO200UpgradeLimit_2018}. This energy resolution is used to define the full-energy region of interest (ROI) for the various modes of $^{124}$Xe decay. For the neutrinoless modes, this corresponds to a narrow energy window around the Q-value.

For the filtering of single energy depositions within an event, the reconstruction can only be based on the S2. To model the broadening of the charge-only energy resolution due to recombination fluctuations, we scale $b$ in the above formula to a value of 4.4. This gives a charge-only resolution of about 6\,$\%$ at $\sim$500~keV, consistent with measurements reported in the literature \cite{Aprile:2011dd,Jewell_2018}.

\subsection{Event reconstruction and efficiency calculation}

This analysis utilizes the information on the various energy depositions at a given spatial position in all three dimensions. In order to reconstruct and validate the efficiency for detecting the unique event topologies, several filtering and clustering steps have to be performed\footnote{We note that the application of machine learning techniques, such as boosted decision trees, in these filtering and clustering steps would likely enhance the signal efficiencies and background rejection. However, their application exceeds the scope of this work, so the conventional analysis outlined here can be regarded as a baseline scenario.}.

First the events are filtered by the total energy deposited in the detector, in order to account for events where decay products leave the detector. This criterion is a fixed value, only broadened by the energy resolution for the neutrinoless modes, but a broad range with a maximum cut-off at the Q-value and a decay-dependent threshold for the two-neutrino decays.

For any remaining event the vertices are sorted by their axial position in the detector (z-coordinate) and these vertices are grouped within a spatial range determined by the assumed position resolution of the detector in the axial direction. For detector configurations where a separation in the radial direction (x-y-coordinate) is also possible, the grouping algorithm also takes separations in x-y into account -- according to the assumed position resolution. The energies of all vertices within each group are summed and provide the individual S2 signals that a detector with the chosen properties would see.

From this point the further filtering targets the reconstruction of the vertices of the annihilation products of the positrons. The procedure is analogously applied to the de-excitation $\upgamma$s in the case of the 0$\upnu$ECEC. It is depicted together with an illustration of the spatial clustering in Figure~\ref{fig:analysis}. All clustered energy depositions of a given event are permuted for each possible interaction combination and the total sum of the energy is compared against the expected value, which is e.g. 511 keV for each $\upgamma$ produced in the positron's annihilation. The combination with the smallest difference between the summed energy and the expected value is then removed from the list of energy depositions if it lies within the energy resolution around the expected value. This raises the counter of measurable signatures by one. Afterwards, this procedure is repeated until all desired signatures have been found and the counter matches the expectation (e.g. 4 in 2$\upnu\upbeta^+\upbeta^+$). For any left-over energy it is then checked if it fulfills the requirement for the point-like deposition expected from the positron and/or the electron capture signal.
In case of 0$\upnu$EC$\upbeta^+$/2$\upnu$EC$\upbeta^+$ a single merged energy deposition of the positron and atomic relaxation processes is expected. While this requirement is a fixed maximum value for a single signature in case of the neutrinoless mode, it is again a continuous distribution ranging from zero or the $31.81\,\text{keV}$ K-shell hole energy to a cut-off depending on the Q-value. The requirement removes energy signatures which are merged by the detector due to the aforementioned limited spatial and time resolution. 
If not all signatures have been found or if the remaining energy is not a single deposition, the event is discarded. The ratio of all events which survive the filtering algorithm and the original generated number of events corresponds to the desired efficiency $\epsilon$. 

\subsection{Influence of thresholds, detector size and position resolution}
\begin{figure}
    \centering
    \includegraphics[width=\linewidth]{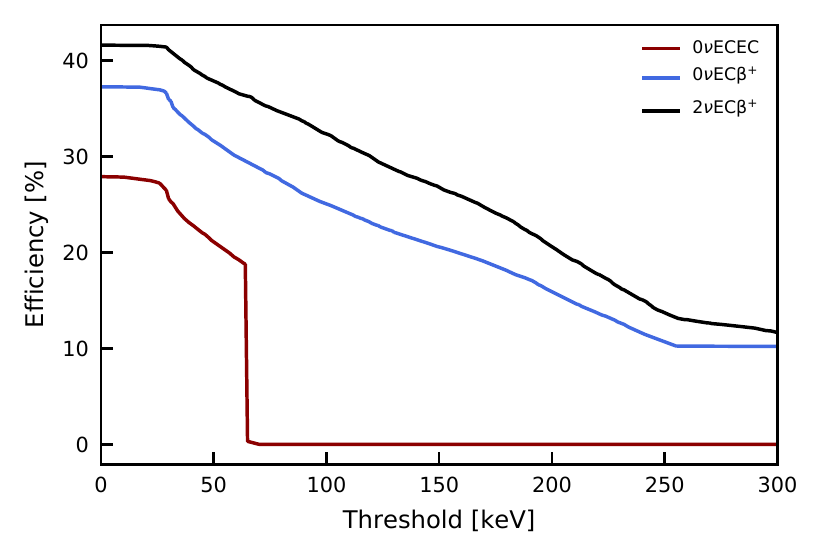}
    \caption{Dependence of the charge-only energy threshold on the detection efficiency for selected decay channels. The efficiency for the 0$\upnu$EC$\upbeta^+$ (blue) and 2$\upnu$EC$\upbeta^+$ (black) show a decrease with energy up to about 250\,keV with efficiencies ranging from about 41\,$\%$ to 10\,$\%$. More striking is the behavior of the 0$\upnu$ECEC (red), which has a sharp cut-off as soon as the double electron capture energy (64.3\,keV) is below the threshold. Since this signature is required within this analysis in order to provide a clear evidence and necessary background suppression, this will automatically drop the efficiency to zero. For this example a position resolution of 10\,mm has been used in both directions.}
            \label{fig:threshold}
\end{figure}

We first investigate the effect of the S2 energy threshold on the detection efficiency. While for Dark Matter detectors the threshold usually is only a few keV thanks to amplification via electroluminescence, the situation is different for an experiment like nEXO, which will measure charge directly. 
In this case, the electronics noise in the readout circuit introduces a larger energy threshold and thus influences the efficiency for the example decay modes as shown in Figure~\ref{fig:threshold}. 
In this work, the threshold is implemented in the simulation simplified as a sharp cut-off for any given energy signature, assuming a position resolution of 1\,cm in radial and axial direction. It is evident that the efficiency depends on this energy threshold. Therefore, an improvement from $\mathcal{O}$(100\,keV) as achieved in EXO-200 \cite{exo-Albert:2013gpz} would be beneficial for nEXO as this has a direct impact on the sensitivity for any given decay channel. Especially, in order to look for a possible smoking gun evidence of the 0$\upnu$ECEC decay, a threshold below the energy of twice the K-shell electron energy is necessary. In the following we assume that a sufficiently low threshold is achieved that it can be considered negligible.

\begin{figure}
    \centering
    \includegraphics[width=\linewidth]{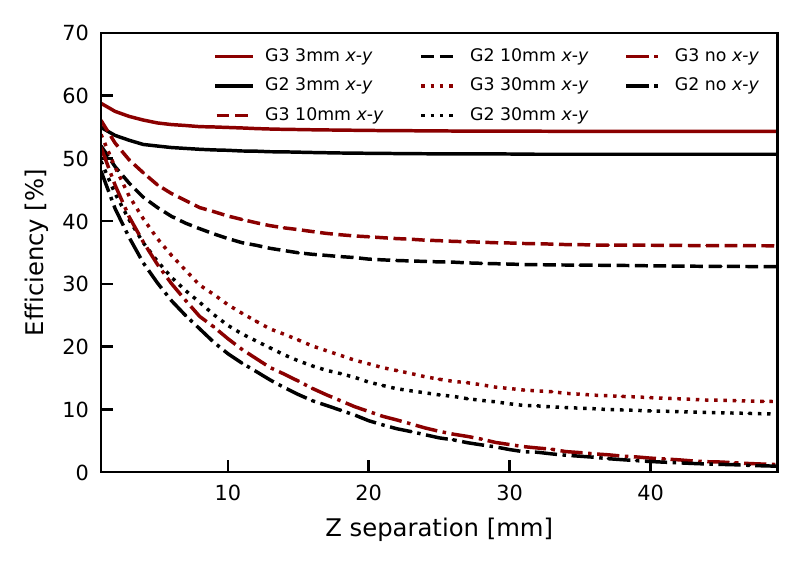}
    \caption{Comparison of efficiencies for the $0\upnu$EC$\upbeta^+$ decay, for different x-y resolutions and detector sizes as a function of the axial resolution. The black (red) lines show the efficiencies for a G2 (G3) detector with $3\,\text{mm}$ (solid), $10\,\text{mm}$ (dashed), $30\,\text{mm}$ (dotted) and no (dashdotted) x-y resolution.}
            \label{fig:eff_results}
\end{figure}

Next, we investigate the effect of a detector's position resolution on the detection efficiency. Our results are shown in Figure~\ref{fig:eff_results} where we emphasize the importance of x-y resolution. For any detector with an axial position resolution (z-coordinate) of a few mm, which is fundamentally limited by electron diffusion, an additional resolution of event topologies in the radial direction is highly beneficial. Already at an achieved 10\,mm separation in the axial direction, an x-y resolution of also 10\,mm can improve the efficiency by more than a factor of two. For a nEXO-type detector this resolution is mostly a function of the pitch of the charge readout strips \cite{nEXO_PCDR_2018}, and therefore can become as small as a few mm. The situation is less clear for dual-phase detectors used in Dark Matter searches; no detector dedicated for Dark Matter search has reported its x-y resolution for multiple energy depositions arriving at the charge detection plane simultaneously. In principle this should be achievable by pattern recognition in the top array of the detector, and is a good candidate for future work in better matching algorithms and machine learning techniques.

Finally, an interesting comparison arises between a nEXO like detector and a G3 Dark Matter experiment, as both could have the same amount of $^{124}$Xe within different-sized detector volumes. The influence of the detector size on the efficiency for the decay mode of 0$\upnu$EC$\upbeta^+$ is shown in Fig.~\ref{fig:eff_results}. It is evident that an increased detector size only increases the efficiency by a few $\%$. This is due to the ratio of events leaving the detector in comparison to the events confined in the full volume. Therefore, the findings for a G3 detector that are summarized in Table \ref{tab:eff_results} are approximately also valid for a nEXO-like detector.
\begin{table*}[]
    \centering
    \begin{center}
            \caption{Efficiencies for all evaluated decay channels in a G2 and a G3 experiment assuming three different radial resolutions and an axial position resolution of 10\,mm. Threshold effects are considered to be negligible.} 
            \label{tab:eff_results}
            \begin{tabular}{c c c c c c c}
             \hline
         \textbf{Decay channel} & \multicolumn{2}{c}{\textbf{Only z [\%]}} & \multicolumn{2}{c}{\textbf{30\,mm x-y [\%]}} & \multicolumn{2}{c}{\textbf{10\,mm x-y [\%]}} \\
         & \textbf{G2} & \textbf{G3} &\textbf{G2} & \textbf{G3} & \textbf{G2} & \textbf{G3}\\
         \hline
         \hline
         $2\upnu\text{EC}\upbeta^+$ & 22 & 24 & 27 &31 & 42 & 47 \\
         $2\upnu\upbeta^+\upbeta^+$ & 4 & 4 & 14 &17 & 31 & 35 \\
         \hline
         $0\upnu\text{EC}\upnu\text{EC}\footnote{Here we considered the most probable branch (57.42$\%$) with a three-fold $\upgamma$-signature. An analysis using the two-fold signatures would yield higher efficiency but can add coincidental $\upgamma$-backgrounds, which would weaken the sensitivity of a given search.}$ & 4 & 5 & 15 & 19 & 28 & 33 \\
         $0\upnu\text{EC}\upbeta^+$ & 19 & 21 & 23 & 27 & 37 & 41 \\
         $0\upnu\upbeta^+\upbeta^+$ & 2 & 2 & 8 & 10 & 25 & 29 \\
        
        \hline
      
            \end{tabular}
    \end{center}
\end{table*}

\section{Backgrounds}
\label{sec:background}
From the above analysis, it is clear that the most experimentally accessible decay channels are the $0\upnu/2\upnu\text{EC}\upbeta^+$. As described, the key feature in a search for $\upbeta^+$-emitting decay modes is the ability to reject backgrounds using the distinct event topology. We consider possible sources of backgrounds below and estimate the expected rates of events passing the topological selection criteria described in Section~\ref{sec:analysis}.

As comparison points, we compute the expected number of $^{124}$Xe decays per tonne-year exposure of $^{\text{nat}}$Xe (corresponding to 0.95~kg-year of $^{124}$Xe) using the half-lives estimated in Table~\ref{tab:half-lives-2nu}, Table~\ref{tab:half-lives_0nu} and Table~\ref{tab:half-lives_relation_0nu}. After including the respective efficiencies for a G2 experiment with 10 mm resolution in x-y-z and assuming a natural xenon ($^\text{nat}$Xe) target, we expect $8.3\pm2.9$~decays per tonne-year for $2\upnu\text{EC}\upbeta^+$. Under the assumption of light-neutrino exchange and given the most optimistic assumptions described above, we expect a rate of less than $2.6\cdot 10^{-2}$~decays per tonne-year for $0\upnu\text{EC}\upbeta^+$.

\subsection{Radiogenic backgrounds from detector materials}

Gamma rays from radioactivity in the laboratory environment and detector construction materials are a primary background in rare event searches. There are two main concerns for the analysis presented here: first, that a $\upgamma$-ray Compton-scatters multiple times and produces the expected event signature. Second, that a $\upgamma$-ray of sufficient energy creates a positron by pair production. In the latter case, the positron will annihilate and produce a background event which, by design, passes our event topology cuts.

We investigate the sources for falsely identified events from the $^{238}$U and $^{232}$Th decay chains, the most common sources of radiogenic backgrounds in most $0\nu\beta\beta$ searches. For each decay step within the chain, 10$^7$ events\footnote{By simulating the same number of events for each decay we implicitly assume decay chain equilibrium. This is not necessarily realized in actual detector construction materials.} have been uniformly generated in a copper shell of 1\,cm thickness surrounding the liquid xenon volume of a G2-sized detector using Geant4. Afterwards, the events which interacted in the active volume were run through the respective event search algorithms for $2\upnu\text{EC}\upbeta^+$ and $0\upnu\text{EC}\upbeta^+$. We find that the only relevant decays are $\upbeta$-decays into excited daughters, as only these produce $\upgamma$s of sufficient energies. 
 
For the neutrinoless case there are two particularly problematic transitions.
The first is the $\upbeta$-decay of $^{214}$Bi in the $^{238}$U-chain, which has a small branching to the 2880\,keV state of $^{214}$Po. If this $\gamma$-ray interacts via pair production, it creates an event identical to our signal directly in the ROI. We find that $1.5\cdot 10^{-6}$ events per $^{214}$Bi primary decay pass the selection criteria. The second problematic transition is the decay of $^{208}$Tl to $^{208}$Pb in the $^{232}$Th-chain, for which there are various transitions in which different $\upgamma$-rays are detected in coincidence with the one from the 2614\,keV state. Such events can deposit enough energy to create events in the ROI, and may similarly produce a sequence of energy depositions which pass our topological criteria.
We find that $1.3\cdot 10^{-4}$ events pass our cuts per $^{208}$Tl primary, but the 35.9\,\% branching fraction for creating $^{208}$Tl in the first place reduces its impact in a real detector to $4.5\cdot 10^{-5}$ events per $^{232}$Th primary decay. Both sources of background can be reduced by a subselection of an inner volume in the active volume of the detector, commonly referred to as a ``fiducial volume cut." As different $\upgamma$-rays with energies below 300\,keV are paired with a high energy $\upgamma$ in the $^{208}$Tl decay signature, fiducializing is especially effective against these events; we find that cutting away the outer 10$\,$cm of LXe reduces its background contribution by almost an order of magnitude. For a 20\,cm cut, no event out of the 10$^7$ simulated for any isotope passes the selection criteria. We conclude that these backgrounds can therefore be eliminated in a real experiment (depending on the actual $^{238}$U/$^{232}$Th contamination) by selecting an appropriate fiducial volume. 

Radiogenic backgrounds have a greater impact on 2$\upnu\text{EC}\upbeta^+$ searches, as the larger energy window allows more events to pass the selection criteria. We find three isotopes in the $^{238}$U chain producing events which pass our selection criteria, with decays of $^{214}$Bi into different excited states of $^{214}$Po being the major background component ($>$99\,\%). The surviving fraction for the total chain is $6.9\cdot 10^{-3}$ events per $^{238}$U primary decay without a fiducial volume selection. This is reduced to $1.5\cdot 10^{-3}$ and $1.1\cdot 10^{-4}$ decays per primary with the 10\,cm and 20\,cm cuts respectively. For the $^{232}$Th-chain, the $^{208}$Pb $\upgamma$-rays following $^{208}$Tl $\upbeta$-decay are again the main contributor ($\sim75\%$). However, $\upgamma$-rays from $^{228}$Th after the $\upbeta$-decay of $^{228}$Ac also contribute ($\sim23\%$), as well as a small contribution ($\sim2\%$) from excited states of $^{212}$Pb following the $\upbeta$-branch of the $^{212}$Bi decay. The surviving fractions for the whole chain are $7.3\cdot 10^{-3}$, $1.5\cdot 10^{-3}$ and $1.3\cdot 10^{-4}$ events per primary $^{232}$Th decay with no fiducial volume cut, a 10\,cm cut and a 20\,cm cut, respectively. Due to the less-stringent energy selection the fiducial volume cuts are less efficient for the $^{208}$Tl events in the two-neutrino case, but still significantly reduce the background contribution.

\begin{figure}[t]
    \centering
    \includegraphics[width=\linewidth]{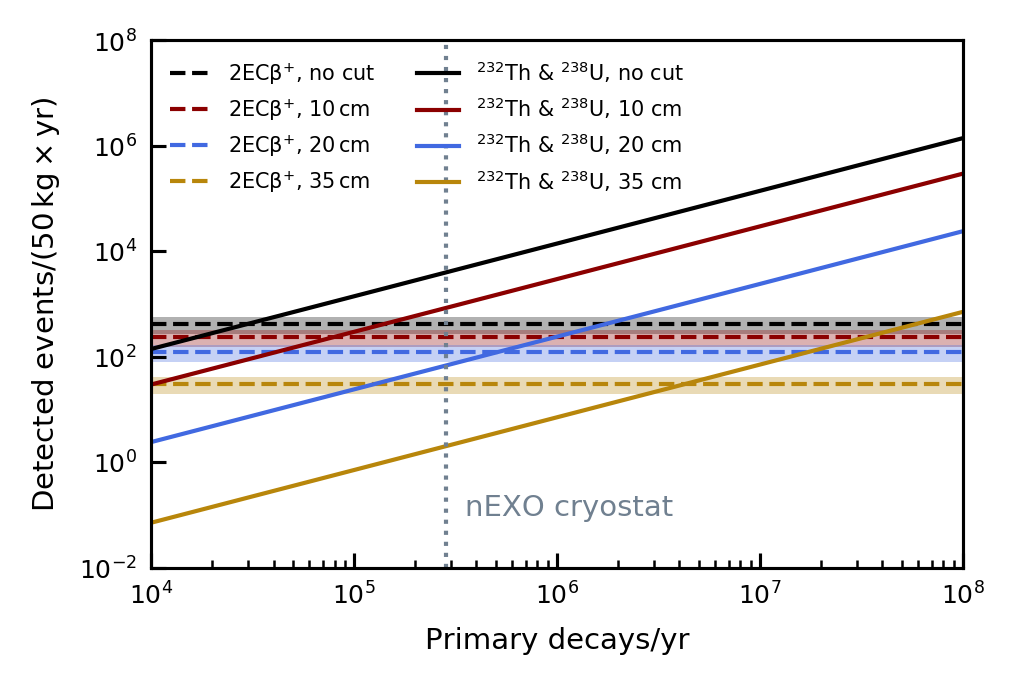}
    \caption{Expected events detected falsely as signal for a given number of primary decays per year for the $^{238}$U and $^{232}$Th decay chains. A reduction from the expectation in the full detector volume (solid black) is achieved by cutting the fiducial volume in all dimensions by 10\,cm (red), 20\,cm (blue) or 35\,cm (gold). For reference the number of expected 2$\upnu\text{EC}\upbeta^+$ signal events is shown for a 50 kg-year \isotope[124]{Xe}-exposure (dashed black) and for the respective fiducial volumes (dashed red, blue and gold) with correspondingly reduced exposures. The expected number of \isotope[232]{Th} and \isotope[238]{U} primary decays per year for the nEXO cryostat is indicated as the dotted gray line.}
    \label{fig:gamma_new}
\end{figure}

In conclusion, two factors play a role for the exact evaluation of a given experimental setting: the fiducial volume cut and the actual amount of contaminants surrounding the TPC. While this study cannot provide an answer for all given experimental settings -- this would need a dedicated Monte Carlo study following a material radioassay -- we use reported contamination levels and experimental details projected for the nEXO experiment (reported in Ref.~\cite{NEXOSensitivity_2018}) to benchmark our calculations. Our approximate evaluation of a nEXO-like experiment is provided in Fig. \ref{fig:gamma_new}. The nEXO experiment identifies the main source of external $\upgamma$-ray backgrounds as the copper cryostat, for which the collaboration reports $^{238}$U and $^{232}$Th concentrations of $0.26\,\text{ppt}$ and $0.13\,\text{ppt}$, respectively. This corresponds to
$2.8\cdot 10^{5}$ primary decays per year as indicated in Figure~\ref{fig:gamma_new} by the dotted gray line. Accordingly, it would only require a mild $10-20$\,cm fiducial volume cut to achieve a favorable signal to background ratio\footnote{This assumes the $\sim650\,\text{kg}$ copper TPC vessel, the main contributor according to \cite{NEXOSensitivity_2018}, as the sole background source.}.
Dark matter experiments, on the other hand, are optimized for the low-energy regime, and typically have higher background levels in the $\sim$MeV regime, and may therefore require more aggressive fiducial cuts to achieve a similar signal-to-background ratio. 

We emphasize that these results are only approximate, and that in a full likelihood analysis the modeling of the events' spatial components and energy distributions will improve the signal to background ratio beyond what has been discussed above. More precise estimates of the impact of these backgrounds are beyond the scope of this work, but will be necessary to understand the true impact of externally-produced $\upgamma$-ray backgrounds in real experiments.

\subsection{$^{222}$Rn}

$^{222 }$Rn may dissolve into the active LXe volume and create backgrounds via $\upbeta$-decays that emit $\upgamma$-rays ($\alpha$-decay events can be easily rejected by the ratio of ionized charge to scintillation light~\cite{EXO200_AlphaIon_2015}). There are only two $\upbeta$-decays in the $^{222}$Rn chain with enough energy to create backgrounds in this analysis. The first, $^{214}$Bi, is accompanied by the subsequent $\alpha$ decay of $^{214}$Po, which occurs with $T_{1/2} =  164\,\upmu$s. Thus, we assume that it can be rejected via a coincidence analysis. The second, $^{210}$Bi, has a Q-value of 1.2\,MeV -- just at the low-energy end of our region of interest for the 2$\upnu$ decays, but well below the ROI for 0$\upnu$ signals -- and decays with no accompanying $\upgamma$. Therefore, it almost always is a single-scatter signal and does not pass our cuts.

\subsection{Charged-current scattering of (anti)neutrinos}
Charged-current (CC) scattering of neutrinos and antineutrinos, while rare, may produce positrons which can exactly mimic our signal of interest.

The CC scattering of low-energy antineutrinos produces a fast positron in the final state.  Here we consider two sources of antineutrinos: nuclear reactors and radioactive decay in the earth (geoneutrinos). Both of these are sources of electron antineutrinos in the few-MeV range. The threshold for the charged-current reaction is set by the mass difference between the xenon isotopes and their iodine isobars. The cross-sections as a function of energy were computed in Ref.~\cite{PirinenChargedCurrentScattering}, and were obtained in tabular form from the authors. 

We calculate the expected rates for geoneutrinos using the two xenon isotopes with the lowest CC reaction thresholds: $^{129}$Xe and $^{131}$Xe, which have thresholds of 1.2$\,$MeV and 2.0$\,$MeV, respectively. 
Convolving the energy spectra and flux with the cross section, we find that the rates for $^{129}$Xe and $^{131}$Xe are 5.0$\times 10^{-8}$ and 4.9$\times 10^{-6}$  events per tonne-year of $^\text{nat}$Xe exposure, respectively. In a G3 detector filled with $^\text{nat}$Xe, there will, therefore, be less than 0.01 events in a 10-year exposure, rendering this background negligible. An experiment using xenon enriched in the heaviest isotopes ($^{134}$Xe and $^{136}$Xe) will be completely insensitive to geoneutrinos due to the high threshold for CC reactions. The flux of geoneutrinos is expected to vary by a factor of $\sim$2 across the globe, so we do not expect these conclusions to depend on the location of the experiment. 

We carry out a similar calculation for reactor antineutrinos, which in contrast are highly location dependent. We assume three possible locations for an experiment: SNOLAB (in Sudbury, Ontario, CA), Sanford Underground Research Facility (in Lead, South Dakota, USA), and Laboratori Nazionali del Gran Sasso (in L'Aquila, Abruzzo, Italy). The reactor antineutrino flux at each site is calculated using reactor power and location data in the Antineutrino Global Map reactor database~\cite{AGM2015}. The antineutrino flux and energy spectra are computed using the empirical models given in Ref.~\cite{Huber2011}. For simplicity, we neglect neutrino oscillations, meaning our expected rates will be overestimated. Of the three candidate locations, the flux is highest at SNOLAB, primarily due to the presence of nearby reactors in Kincardine and Pickering, ON. In this case, we calculate an expected CC scattering rate of $9.1\times10^{-7}$ and $3.6\times10^{-6}$ events per tonne-year for scattering on $^{129}$Xe and $^{131}$Xe, respectively. The expected rates at the other candidate locations are smaller by at least an order of magnitude.

In contrast to antineutrinos, CC scattering of low-energy neutrinos does not directly create positrons. There are, however, two possible backgrounds that may arise from this reaction: the emission of a fast electron and a daughter nucleus in an excited state (which can de-excite and create additional energy deposits that may mimic the signal event topology), and the creation of a daughter radioisotope which later decays via $\upbeta^+$-emission. 

The primary source of neutrinos incident on deep underground detectors is the sun. For our purposes, the most important are those produced by the decay of $^{8}$B, which have energies of $\sim1-10\,$MeV. These are the only solar neutrinos with enough energy to react above threshold \emph{and} populate an excited state in the daughter nucleus, for all xenon isotopes. The energy-averaged cross-section for these reactions is tabulated in Ref.~\cite{PirinenChargedCurrentScattering}, and is of $\mathcal{O}\left(10^{-42}-10^{-41}\right)\,$cm$^2$. This may produce 10's of events per tonne-year for each isotope in a $^{\text{nat}}$Xe detector. However, scattering into low-lying excited states is suppressed, with partial cross-sections an order of magnitude smaller than the total. Therefore, most of the events will deposit too much energy in the detector and will be rejected. 
We find that the low probability of the remaining events passing our topological selection criteria renders these backgrounds negligible for both the $2\upnu\text{EC}\upbeta^+$ and the $0\upnu\text{EC}\upbeta^+$ decay modes.

Neutrino CC scattering on xenon may create radioactive isotopes of caesium in the liquid target. Of particular concern are $^{128}$Cs and $^{130}$Cs, which each have half-lives of $<1\,$hr and can decay via $\upbeta^+$-emission with Q-values of 3.9\,MeV and 2.9\,MeV, respectively, exactly mimicking our expected event signature. Again using the $^8$B-averaged cross sections from Ref.~\cite{PirinenChargedCurrentScattering}, we calculate a production rate of 0.02 nuclei of $^{128}$Cs and 0.07 nuclei of $^{130}$Cs per tonne-year of $^\text{nat}$Xe exposure. The resulting $\upbeta^+$ decays are distributed across a broad spectrum, and our simulations indicate that they will be a small background for the $2\upnu$EC$\upbeta^+$ process, with expected rates an order of magnitude lower than the expected signal rate. The narrow ROI for 0$\upnu$ searches will render these backgrounds negligible. There are also two isotopes of xenon with CC reaction thresholds low enough to react with CNO, $^{7}$Be, and $pp$ neutrinos: $^{131}$Xe and $^{136}$Xe. However, the relevant Cs daughter isotopes have half-lives of $\mathcal{O}$(10) days. Next-generation experiments plan to recirculate and purify the liquid xenon with a turnover time of $\sim$2 days~\cite{nEXO_PCDR_2018}, meaning these isotopes will likely be removed from the detector before they decay.

\subsection{Neutron-induced backgrounds}

A final possibility for backgrounds are those from neutron scattering or capture. In neutron capture, the daughter nucleus is generally left in a highly-excited state, and relaxes to the ground state via the emission of several $\upgamma$-rays. As the sum total of the energy lost in this process is well above the Q-value for $^{124}$Xe decay, we expect these events will be easy to reject and we neglect this as a background source.

For neutron scattering, it is of particular interest to estimate the activation rate of Xe-radioisotopes that may decay via $\upbeta^+$ emission in the region of interest. We identify the fast neutron scattering $^{124}$Xe($n,2n$)$^{123}$Xe reaction as the only one of significance. It has a neutron-energy threshold of 10.5$\,$MeV and the cross-section reaches $\sim$1.4~barn at a neutron energy of $\sim20\,$MeV~\cite{ENDFB_VIII0}. The high threshold prevents radiogenic neutrons (which come from ($\alpha$,n) reactions in the laboratory environment) from producing this background, but muon-induced neutrons, which can extend in energy up to the GeV scale, are of concern. We use an estimate of the muon-induced neutron flux at Gran Sasso of $10^{-9}\,$n/cm$^2$/s and multiply by a factor of 10$^{-2}$ to account for the expected reduction due to shielding typically employed in these experiments~\cite{XENON1T_WaterShield_2014}. We find an expected activation rate of $\sim$10$^{-3}$ atoms per kg ($^{124}$Xe) per year, each of which we assume will produce a background event in the TPC. However, this decay has a small branching ratio for $\upbeta^+$ decay, and even then always proceeds to an excited state of $^{123}$I. Accordingly, it is unlikely to pass our selection criteria, and we consider it negligible.

\subsection{Summary}

After considering an exhaustive list of background sources, for $2\upnu\text{EC}\upbeta^+$ we conclude that the only significant background originates from external $\upgamma$-rays. With strong fiducial volume cuts, a likelihood-analysis utilizing energy information and $\upgamma$-background suppression, near-future G2 Dark Matter experiments have a strong chance of measuring this decay mode. For $0\upnu\text{EC}\upbeta^+$, we conclude that the searches in G2 and G3 experiments will basically be ``background-free," and the sensitivity will only be limited by the detection efficiencies and the attainable $^{124}$Xe exposure in each experiment.

\section{Sensitivity}
\label{sec:sensitivity}
The half-life measured by a detector configuration with no expected background for a number of $N$ observed decay events is given by
\begin{equation}
    T_{1/2} = \frac{\ln(2) N_\text{A} \times \epsilon \times \text{m} \times t}{N \times M_{\text{Xe}}}.
\end{equation}
Here, $N_\text{A}$ is Avogadro's constant, $\epsilon$ is the detection efficiency, and $M_{\text{Xe}}$ corresponds to the molar mass of $^{124}$Xe. The available mass of $^{124}$Xe, $m$, and the measurement time, $t$, depend on the detector configuration. If no events are observed and if a Poissonian process without background is assumed, a $90\,\%\,\text{C.L.}$ lower limit on $T_{1/2}$ can be calculated by inserting $N = 2.3$.

For a detector with 10\,mm position resolution in the axial as well as the x-y direction, the expected half-life can be calculated as a function of exposure using the previously calculated efficiencies. The sensitivities for a G3 experiment with a 500\,kg-year exposure are summarized in Table \ref{tab:results_half_lives} for all decay modes. A similar exposure would be possible in a G2 detector enriched to 50\,kg of $^{124}$Xe; the only difference is the $\sim 10\%$ decrease in detection efficiency due to the increased probability of energy being deposited outside the sensitive volume of the detector. The sensitivities are compared to the range of theoretical predictions from Table \ref{tab:half-lives-2nu} for $2\upnu$-decays, and Tables \ref{tab:half-lives_0nu} and \ref{tab:half-lives_relation_0nu} for $0\upnu$-decays. 

Regarding the two-neutrino decays 2$\upnu\text{EC}\upbeta^+$ will likely be detected by a G3 experiment, but are already be accessible to a G2 detector with a $^\text{nat}$Xe target if the $\gamma$-background is properly addressed. However, due to an unfavourable phase-space 2$\upnu\upbeta^+\upbeta^+$ will likely be out of reach of even a G3 detector. On the neutrinoless side 0$\upnu\upbeta^+\upbeta^+$ is also pushed to experimentally inaccessible half-lives by the unfavourable phase-space. An eventual detection of $0\upnu\text{ECEC}$ relies on the presence of a sufficient resonance enhancement that could boost the decay rate approximately four orders of magnitude. However, given current measurements of decay energies and \isotope[124]{Te} energy levels this is not present \cite{dec-Nesterenko:2012xp, KATAKURA20081655}. A final independent measurement is recommended by the authors of \cite{dec-Nesterenko:2012xp} would be needed for a final verdict on the detection prospects of this decay. Thus, it is evident that the most promising neutrinoless decay is 0$\upnu\text{EC}\upbeta^+$. 

\begin{figure}[t]
    \centering
    \includegraphics[width=\linewidth]{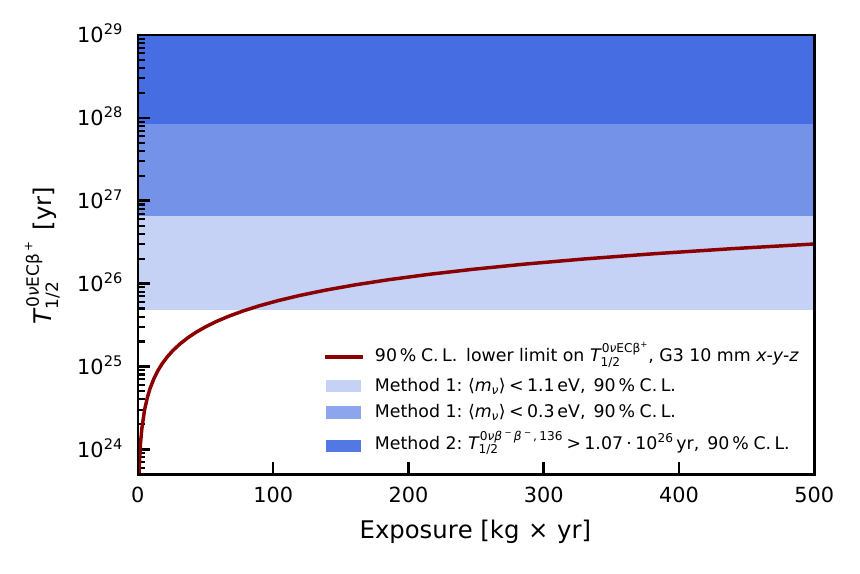}
    \caption{Projected $90\,\%\text{C.L.}$ lower limit on $T_{1/2}^{0\upnu\text{EC}\upbeta^+}$ for a background-free experiment with 10 mm resolution in x-y-z, as a function of the exposure (red). This calculation assumes the G3 geometry; the sensitivity curve decreases by $\sim$10\% for a G2-sized detector at all exposures. Three ranges of lower limits on the 0$\upnu$EC$\upbeta^+$-decay half-life are shown: the direct calculation (Table \ref{tab:half-lives_0nu}) with $\langle m_\upnu \rangle < 1.1 ,\text{eV}$ (light blue) and with $\langle m_\upnu \rangle < 0.3\,\text{eV}$ (medium blue), and the NME comparison (Table \ref{tab:half-lives_relation_0nu}) using the KamLAND Zen \isotope[136]{Xe} $0\upnu\upbeta^-\upbeta^-$ half-life limit (dark blue). For the direct method the lower bound is given by the weakest limit among the three NMEs for each $\langle m_\upnu\rangle$.}
            \label{fig:results}
\end{figure}

For this decay we compare the experimental sensitivity derived in this study with three possible theoretical scenarios. Scenarios one and two are based on the direct calculation (Method 1, Eq.~\ref{eq:0vdecayrate} and Eq.~\ref{eq:0vececdecayrate}) using the effective neutrino mass range from Eq. \ref{eq:mnu_eff}. Scenario three is based on the comparison of \isotope[124]{Xe}-NMEs with the NME for $0\upnu\upbeta^-\upbeta^-$ of \isotope[136]{Xe} using the KamLAND Zen half-life limit (Method 2, Eq.~\ref{eq:half-life_relation_0nu}). The results are shown as a function of exposure in Fig.~\ref{fig:results}. Within a 500\,kg-year-exposure, a background-free experiment would cover a significant portion of the parameter space given by the KATRIN limit translated to $\langle m_\upnu \rangle$. Once this value is reduced, e.g. by phase cancellations in the PMNS-matrix, the lower limits on the half-life are an order of magnitude above the experimental sensitivity. Assuming the same decay mechanism for \isotope[136]{Xe} and \isotope[124]{Xe} -- here light-neutrino exchange -- the expected half-lives are two orders of magnitude above the experimental sensitivity taking into account the current limits placed by KamLAND Zen. Exposures larger than  $10^4$\,kg-year would be needed to probe this parameter space.
\begin{table*}[t]
    \centering
    \begin{center}
            \caption{Results and theoretical predictions for the various decay channels of \isotope[124]{Xe}. The experimental sensitivity is calculated for a 500 kg-year exposure assuming a G3-experiment with 10\,mm position resolution in all three dimensions, a negligible threshold, and no backgrounds. The range of theoretical predictions for neutrinoless decays is given between the weakest limit from the direct calculation with $\langle m_\upnu \rangle < 1.1\,\text{eV}$ (Table \ref{tab:half-lives_0nu}) and the strongest limit from the comparison with KamLAND Zen (Table \ref{tab:half-lives_relation_0nu}).} 
            \label{tab:results_half_lives}
            \begin{tabular}{c c c c}
             \hline
         \textbf{Decay} & \textbf{Exp. Sensitivity [10$^{26}$ yr]} & \textbf{ Experiment/Theory }\\
          \hline
        \hline
         $2\upnu\text{EC}\upbeta^+$ & 3.3 & $(1.9\pm 0.7) \cdot 10^3$  \\ 
         $2\upnu\upbeta^+\upbeta^+$ & 2.5 & $(1.3\pm 0.4) \cdot 10^{-2}$\\
         \hline
         $0\upnu\text{ECEC}$ & 2.4 & $1.3\cdot 10^{-3}- 1.1\cdot10^{-6}$ \\ 
         $0\upnu\text{EC}\upbeta^+$ & 2.9 & $6.0 - 3.5\cdot10^{-2}$\\ 
         $0\upnu\upbeta^+\upbeta^+$ & 2.4 & $0.3-1.6\cdot10^{-3}$ \\
        
        \hline
     
            \end{tabular}
    \end{center}
\end{table*}

\section{Discussion}
\label{sec:discussion}
This work has summarized the possible decay modes of $^{124}$Xe and investigated possible efficiencies of future liquid xenon detectors to the respective channels. For a G2 Dark Matter detector a detection of 2$\upnu$EC$\upbeta^+$ is feasible given a proper treatment of potential $\upgamma$-backgrounds. An experiment with the expected background of a double-$\upbeta$ decay detector like nEXO, would be able to clearly detect the decay and could study it with precision, if $^{124}$Xe would be added to the xenon inventory. A G3 Dark Matter experiment like DARWIN would have the signal strength to detect this decay with a few thousand signals, but would need to optimize its fiducial volume in order to reduce the $\upgamma$-background. 

For a possible neutrinoless mode of this decay, achieving a background-free experiment is a realistic prospect owed to the decay signatures. However, we have shown that in this case a detection is only within reach of a G3 or an enriched nEXO-like detector for the most conservative half-life predictions. It has to be emphasized that such a scenario would require a mechanism that leads to a difference in the decay of proton-rich nuclei compared to their neutron-rich counterparts. Otherwise it would be excluded by KamLAND Zen.

As mentioned previously, such a mechanism would be an exciting prospect in searches for neutrinoless decays of proton-rich isotopes. If detected, it would provide complementary information on the physical mechanism mediating the decay process. One example for this possibility was studied in detail in Ref.~\cite{dec-Hirsch1994} in the context of left-right symmetric models, in which one assumes that there is a right-handed weak sector in addition to left-handed neutrinos, which can mediate neutrinoless double-$\upbeta$ decays. Detectors with the capability of measuring both isotopes simultaneously may therefore be attractive for both the discovery of the neutrinoless process and subsequent study of the underlying physics.
\begin{figure}[t]
    \centering
    \includegraphics[width=0.5\textwidth]{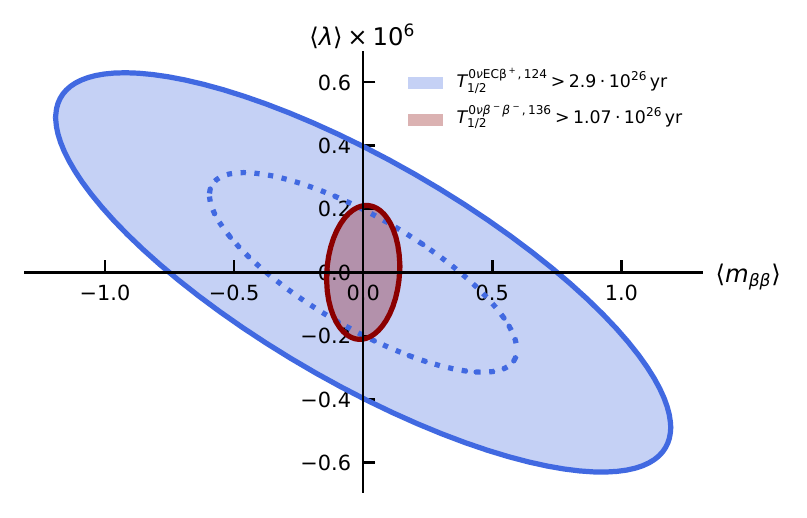}
    \caption{Comparison of exclusion limits at $90\,\%\text{C.L.}$ for left-right symmetric models, in the $\langle m_\upnu \rangle$ vs. $\langle \lambda \rangle$ plane. Parameter space outside the colored regions is excluded. Here we assume $\langle \eta \rangle = 0$. The exclusion limits compare the present limits on the $0\upnu\upbeta^-\upbeta^-$-decay of \isotope[136]{Xe} \cite{KAMLANDZen_2016} with the possible limits on $0\upnu\text{EC}\upbeta^+$ derived in this work. We assume the full 500\,kg-year exposure for the $^{124}$Xe search -- comparable to the 504\,kg-year exposure used for the $^{136}$Xe measurements. The dashed line represents the boundary of the excluded zone after arbitrarily scaling the NMEs for $^{124}$Xe by a factor of two, to mimic uncertainties in NME calculations.}
    \label{fig:RHC_exclusion}
\end{figure}
Here we briefly reexamine the analysis of left-right symmetric models using the projected sensitivities described in this work. By adding right-handed terms to the Standard Model Lagrangian, one derives a new expression for the half-life of neutrinoless second-order weak decays:
\begin{align}
\begin{split}
[T_{1/2}^{\alpha}(0^+_i&\rightarrow0^+_f)]^{-1} = \\
&C_{mm}^\alpha\left(\frac{\langle m_{\upnu} \rangle}{m_e}\right)^2 + 
C_{\eta\eta}^\alpha \langle \eta \rangle^2 + 
C_{\lambda\lambda}^\alpha \langle \lambda \rangle^2 + \\
&C_{m\eta}^\alpha \frac{\langle m_{\upnu} \rangle}{m_e} \langle \eta \rangle +
C_{m\lambda}^\alpha \frac{\langle m_{\upnu} \rangle}{m_e} \langle \lambda \rangle +
C_{\eta\lambda}^\alpha \langle \eta \rangle \langle \lambda \rangle,
\end{split}
\end{align}
where $\alpha$ represents the decay mode ($0\upnu\upbeta^{-}\upbeta^{-}$, $0\upnu \text{EC} \upbeta^{+}$, etc.), $\langle m_\upnu \rangle$ is the effective light neutrino mass defined above, and $\langle \eta \rangle$ and $\langle \lambda \rangle$ are the effective coupling parameters for the new interaction terms containing right-handed currents. The coefficients $C^{\alpha}_{ij}$ are combinations of nuclear matrix elements and phase space factors, and differ between the decay modes. In particular, it was pointed out in Ref.~\cite{dec-Hirsch1994} that the $\lambda$ terms are significantly enhanced in the case of the mixed-mode decays, meaning the shape of the parameter space explored by $0\upnu\text{EC}\upbeta^+$ searches differs from that explored by the more common $0\upnu\upbeta^-\upbeta^-$ experiments. We illustrate this in Figure~\ref{fig:RHC_exclusion}, where we compare the possible limits for $0\upnu\text{EC}\upbeta^+$ derived in this work with the current limits for the $0\upnu\upbeta^-\upbeta^-$ of $^{136}$Xe decay from the Kamland-Zen experiment~\cite{KAMLANDZen_2016}.

We see that the sensitivity of the mixed-mode $^{124}$Xe decay to the effective neutrino mass is significantly weaker; this is due to the reduced phase space in the positron-emitting decay mode. However, the sensitivity of the mixed-mode decay is within a factor of two for the right-handed coupling $\langle \lambda \rangle$, which is within the uncertainties typically assumed for nuclear matrix element calculations (usually a factor of $\sim 3$). Consequently, such a measurement would provide complementary information in the event of a discovery of a 0$\upnu$ decay mode in either isotope. It must be acknowledged that future experiments expect to reach sensitivities considerably larger than the existing limits. Unless the $0\upnu\upbeta^-\upbeta^-$ decay of $^{136}$Xe is just beyond the reach of present experiments, we show that the $^{124}$Xe mixed-mode decays will not be competitive in constraining left-right symmetric models with a G3 experiment's exposure. However, exploring proton-rich isotopes may still provide complementary information in determining the mechanism of lepton number violation; for example, an (unexpected) discovery of neutrinoless decays in either only $^{124}$Xe or in \emph{both}  $^{124}$Xe and $^{136}$Xe could prove that neither the light neutrino exchange nor right-handed currents mediate the decay processes, and could point towards alternative new physics. Therefore, we emphasize that future xenon-based TPC experiments should explore this decay channel, as the striking multiple coincidence structure is straightforward to look for and distinguish from backgrounds. Also the consideration to expand an existing program like nEXO, which would require an additional enrichment on the light mass side after the initial enrichment, could be thought of, in order to gain further insight into the neutrinoless decay modes -- especially once it has been found in $^{136}$Xe.


\begin{acknowledgments}
We thank Javier Men\'endez for calculating the nuclear shell model matrix elements as well as for consultation. We thank Lutz Althueser for providing the XeSim software package, Pekka Pirinen for providing the cross section data used to compute CC solar neutrino backgrounds, and Michael Jewell for useful discussions. 

This work was supported, in part, by DOE-NP grant DE-SC0017970. B.L. acknowledges the support of a Karl Van Bibber fellowship from Stanford University. C. Wi. acknowledges support by DFG through the Research Training Group 2149 "Strong and weak interactions -- from hadrons to dark matter".
\end{acknowledgments}

\bibliographystyle{apsrev}
\bibliography{biblio}

\end{document}